\documentclass{tlp}

\usepackage{latexsym}
\usepackage{bbm}
\usepackage{url}

\newtheorem{proposition}{Proposition}[section]
\newtheorem{lemma}[proposition]{Lemma}

\newtheorem{theorem}[proposition]{Theorem}

\newtheorem{definition}[proposition]{Definition}
\newtheorem{example}[proposition]{Example}

\newcommand{\dom}{\mathrm{dom}}
\newcommand{\false}{\mathbf{false}} \newcommand{\true}{\mathbf{true}}

\newcommand{\ext}[1]{\lceil {#1} \rceil}   \newcommand{\syns}[2]{[\![ {#1} ]\!]_{{\cal S}, {#2}}}
\newcommand{\sem}[1]{\langle \! \langle {#1} \rangle \! \rangle}
\newcommand{\semi}[2]{\sem{#1}_{#2}}
\newcommand{\pfsem}[2]{\semi{#1}{#2}^*}
\newcommand{\switch}[2]{
  \begin{array}{c}
    {#1} \\[-4pt] {#2}
  \end{array}
  }

\newcommand{\Epsilon}{{\cal E}}

\newcommand{\Array}[2]{{#2}[{#1}]}
\newcommand{\Read}[3]{\mathbf{Read}({#1}, {#2}, {#3})}
\newcommand{\Write}[3]{\mathbf{Write}({#1}, {#2}, {#3})}
\newcommand{\subst}[3]{{#1} [ ^{#2} / _{#3} ]}

\newcommand{\TERMS}{{\it TERMS}}
\newcommand{\implies}{\Rightarrow}
\renewcommand{\iff}{\Leftrightarrow}
\newcommand{\pstep}[1]{\quad \{ \mbox{ #1 } \}}
\newcommand{\existsnext}{{\exists \! \, \bigcirc}}
\newcommand{\forallnext}{{\forall \! \, \bigcirc}}

\newcommand{\code}[1]{{\tt #1}}

\submitted{18 March 2002}
\revised{-} 
\accepted{2 July 2003}

\begin{document}
\sloppy

\title[On model checking data-independent systems with arrays
without reset]{On model checking data-independent \\ systems with arrays
without reset \footnote{This work was funded in part by the EPSRC standard
research grant `Exploiting data independence', GR/M32900.
The first author is affiliated to the Mathematical Institute,
Belgrade, and was supported partly by a grant from the Intel Corporation,
a Junior Research Fellowship from
Christ Church, Oxford, and previously by a scholarship from
Hajrija \& Boris Vukobrat and Copechim France SA.
The second author was funded in part by QinetiQ Malvern.
The third author was funded in part by the US ONR.}
}
\author[R.S. Lazi\'c, T.C. Newcomb, and
A.W. Roscoe]{R.S. LAZI\'C\\
  Department of Computer
  Science, University of Warwick, \\ Coventry, CV4 7AL, UK\\
  \email{ranko.lazic@dcs.warwick.ac.uk}
 \and T.C. NEWCOMB and
A.W. ROSCOE\\
Oxford University Computing Laboratory, \\ Wolfson Buildings,
Parks Road, Oxford, OX1 3QD, UK\\
\email{tom.newcomb@comlab.ox.ac.uk} \\
\email{bill.roscoe@comlab.ox.ac.uk}}

\maketitle

\begin{abstract}
   
  A system is data-independent with respect to a data type $X$ iff the
  operations it can perform on values of type $X$ are restricted to
  just equality testing.  The system may also store, input and output
  values of type $X$.
  
  We study model checking of systems which are data-independent with
  respect to two distinct type variables $X$ and $Y$, and may in
  addition use arrays with indices from $X$ and values from $Y$.  Our
  main interest is the following {\em parameterised model-checking}
  problem: whether a given program satisfies a given temporal-logic
  formula for all non-empty finite instances of $X$ and $Y$.
  
  Initially, we consider instead the abstraction where $X$ and $Y$ are
  infinite and where partial functions with finite domains are used to
  model arrays.  Using a translation to data-independent systems
  without arrays, we show that the $\mu$-calculus model-checking
  problem is decidable for these systems.
  
  From this result, we can deduce properties of all systems with
  finite instances of $X$ and $Y$. We show that there is a procedure
  for the above parameterised model-checking problem of the universal
  fragment of the $\mu$-calculus, such that it always terminates but
  may give false negatives.  We also deduce that the parameterised
  model-checking problem of the universal disjunction-free fragment of
  the $\mu$-calculus is decidable.
  
  Practical motivations for model checking data-independent systems
  with arrays include verification of memory and cache systems, where
  $X$ is the type of memory addresses, and $Y$ the type of storable
  values. As an example we verify a fault-tolerant memory interface
  over a set of unreliable memories.
\end{abstract}


Submitted:  March 18, 2002;   accepted: July 2, 2003.

\section{Introduction}

A program is data-independent \cite{Wol86,LN00} with respect to a
data type $X$ if it can only input, output, and assign values of type
$X$, as well as test pairs of such values for equality.  The program
cannot apply any other operation to values of type $X$.

Data-independent programs are common.  Communication protocols are
data-independent with respect to the type that is being communicated.
Nodes of a network protocol may be data-independent with respect to
the type of node identifiers.

Given a program $\cal P$ which is data-independent with respect to a
type $X$, the type $X$ can be seen as a type variable, i.e.\ as a
parameter of $\cal P$, in the sense that it can be instantiated by any
set.  Given a property $\varphi$ in temporal logic, the {\em
  parameterised model-checking problem} asks whether $\cal P$
satisfies $\varphi$ for all instances of $X$.  A variety of
decidability results are known for this and related problems (e.g.\ 
\cite{Wol86,HDB97,LN00,FS01}).

In this paper, we consider programs which are data-independent with
respect to two types $X$ and $Y$, but which can in addition use arrays
indexed by $X$ and storing values of type $Y$.  We focus on the case
where the programs may use the operations for reading and writing
an array component, but where array reset (i.e.\ assigning a given value
of type $Y$ to all array components) is not available.

The techniques which were used to establish decidability of
parameterised model checking for data-independent programs cannot be
used when data independence is extended by arrays.  An array is
indexed by the whole of the type $X$, and it therefore may contain an
unbounded number of values of type $Y$. These values may have been
fixed by previous actions, and although they are not all accessible in
the current state, they may become accessible if their indices appear
in variables of type $X$ in subsequent states.

One motivation for considering data-independent programs with arrays
is cache-coherence protocols \cite{AG96}, more precisely the problem
of verifying that a memory system satisfies a memory model such as
sequential consistency \cite{HQR99}.  Cache-coherence protocols are
data independent with respect to the types of memory addresses and
data values.

Another application area is parameterised verification of network
protocols by induction, where each node of the network is
data-independent with respect to the type of node identities
\cite{CR00}.  Arrays arise when each node is data-independent with
respect to another type, and it stores values of that type.

Given a data-independent program $\cal P$ with arrays and a
temporal-logic formula $\varphi$ referring to control states of $\cal
P$, the main question of interest is whether $\cal P$ satisfies
$\varphi$ for all non-empty finite instances of $X$ and $Y$.

In order to study decidability of this parameterised model-checking
problem, we first consider the abstraction where $X$ and $Y$ are
instantiated to infinite sets, and where arrays are modelled by
partial functions with finite domains.  An undefined array component
represents nondeterminism which is still to be resolved.

We describe a translation of such a program to a
bisimulation-equivalent data-independent program without arrays; it
follows that the $\mu$-calculus model checking problem is decidable in
this case \cite{BCG88,NK00}. The $\mu$-calculus is a branching-time
logic, more expressive than CTL or CTL$^*$ \cite{AH98}.

For a program $\cal P$, any transition system generated by $\cal P$
with finite instances of $X$ and $Y$ is simulated by the transition
system generated by $\cal P$ with infinite instances of $X$ and $Y$.
It follows that there is a procedure for the parameterised
model-checking problem of the universal fragment of the
$\mu$-calculus, such that it always terminates, but may give false
negatives. This fragment of the $\mu$-calculus is more expressive than
linear-time temporal logic.

We also deduce that the parameterised model-checking problem of the
universal disjunction-free fragment of the $\mu$-calculus is
decidable. This fragment of the $\mu$-calculus is more expressive than
reachability, although less expressive than linear-time temporal logic
\cite{HM00}. It can be used to express properties such as ``the system
produces an output every ten time units.'' Such a property could be
checked less naturally using reachability on a modified version of the
system.

As an example, we model a simple fault-tolerant interface working over
a set of unreliable memories. The parameterised model-checking
procedure presented here is used to verify its correctness with
respect to the specification ``a read at an address always returns the
value of the last write to that address until a particular number of
faults occur,'' independently of the size of the memory and of the
type of storable data values.  This program illustrates how our
procedure works, and is a simple representative from the class of
programs to which this paper applies.  More concretely, using our
results it is possible to model and verify some types of
fault-tolerant fully-associative cache systems \cite{PH94},
independently of cache size, memory size, the type of data values, and
page replacement policies.

Our results might be compared to \cite{HIB97}, where it is shown that
data-independent programs with one array, without reset, with infinite
instances of $X$ and $Y$, and with a slightly different modelling of
arrays by partial functions, have finite trace-equivalence quotients.
The parameterised model-checking problem is not considered.  We have
extended this result to allow many arrays, and have shown that model
checking of the $\mu$-calculus is decidable in the infinite-arrays
case, which is a stronger logic than the linear-time temporal-logic
induced by finite trace-equivalence quotients. Also, the parameterised
model-checking problem for finite arrays is not considered in
\cite{HIB97}, whereas we have developed decidability results for these
systems.

This paper clarifies a technique described in \cite{McM99}, which
promotes the use of abstract interpretation for programs with arrays.
The programs considered there are more general than ours as the arrays
may be multi-dimensional and of varying index and data types.
Temporal case splitting is used to consider only a finite portion of
the arrays; at the other locations a read operation returns a special
symbol $\bot$ which represents any element in the type.  Datatype
reduction, a standard abstraction used for data-independent programs
\cite{ID96}, is then used to deal with the remaining values stored in
the arrays.  This is a similar strategy to that used in the proofs in
this paper, although \cite{McM99} presents no decidability results
about the technique apart from stating that the problem is undecidable
in general. We have identified a smaller, yet still interesting class
of programs and shown that there is an automatic parameterised
model-checking procedure for them.

An advantage of this paper over both these related works is that we
use a syntactic transformation to remove the arrays. This admits the
application of orthogonal state reduction techniques, such as further
program transformations or advanced model checking algorithms, eg.\ 
using BDDs \cite{BC+92}.

The contributions of this paper are as follows. We describe an
automatic procedure for model checking a programming language useful
for prototyping memory systems such as caches. We extend the result
about infinite arrays in \cite{HIB97}, and also show how our result
relates to questions about finite arrays. This allows us to prove
properties about parameterised systems: for example, that memory
systems can be verified independently of memory size and data values.
We also identify a subclass of the programs considered in \cite{McM99}
and prove the decidability of model checking them.  Decidability
results are important because they provide verification procedures
which are guaranteed to terminate for every instance of the problem,
with a correct answer.

The rest of this paper is organised as follows. Section
\ref{sect:prelims} introduces some standard definitions and
preliminary results, and then the language of programs we will be
considering is defined in Section \ref{sect:lang}. Section
\ref{sect:infinite} considers the case that the types $X$ and $Y$ are
infinite, and from this we deduce results about all the cases when
they are finite in Section \ref{sect:finite}. We conclude with a
summary and discussion of future work in Section
\ref{sect:conclusion}.

\section{Preliminaries}

\label{sect:prelims}

In this section we introduce transition systems as our modelling
language, and our language of specifications, the modal
$\mu$-calculus.

\subsection{Transition systems}

\begin{definition}
A {\em transition system} is a structure $(Q, \delta, \ext{\cdot},
P)$:
\begin{itemize}
\item $Q$ is the {\em state space},
\item $\delta : Q \rightarrow 2^Q$ is the {\em successor function},
  giving the set of possible next states after the given state,
\item $P$ is a finite set of {\em observables},
\item $\ext{\cdot} : P \rightarrow 2^Q$ is the {\em extensions function}.
\end{itemize}
\end{definition}
Thus $\ext{p}$ is the set of states in $Q$ that have some observable
property $p$. In this paper, $p$ will typically be a boolean variable
of the program under consideration, and will be observed at exactly
the states where the value of the variable is ``true''.

\begin{definition}
A {\em trace} $\pi$ of a transition system is a finite sequence of
observables $p_1 p_2 ... p_l$ such that there exists a sequence of
states $s_1 s_2 ... s_l$ from $Q$ where $s_{i+1} \in \delta(s_i)$
(for $i = 1 ... l-1$) and $s_i \in \ext{p_i}$ (for $i = 1 ... l$).
We will write $\pi(i)$ to mean $p_i$, the $i$th observable in the
trace $\pi$.
\end{definition}

Given two transition systems ${\cal S}_1 = (Q_1, \delta_1,
\ext{\cdot}_1, P)$ and ${\cal S}_2 = (Q_2, \delta_2, \ext{\cdot}_2,
P)$ over the same observables $P$, it is possible to compare them in
the following ways.

\begin{definition}
A relation $\preceq\ \subseteq Q_1 \times Q_2$ is a {\em simulation} if 
$s \preceq t$ implies the following two conditions:
\begin{itemize}
\item[1.] For all observables $p$, $s \in \ext{p}_1$ if and only if $t \in
  \ext{p}_2$.
\item[2.] For each state $s' \in \delta_1(s)$, there is a state $t' \in
  \delta_2(t)$ such that $s' \preceq t'$.
\end{itemize}
\end{definition}

\begin{definition}
  A relation $\approx\ \subseteq Q_1 \times Q_2$ is a {\em bisimulation} if 
  it is a simulation and $s \approx t$ also implies the following condition:
\begin{itemize}
\item[3.] For each state $t' \in \delta_2(t)$, there is a state $s' \in
  \delta_1(s)$ such that $s' \approx t'$.
\end{itemize}
\end{definition}

\subsection{The $\mu$-calculus}

The following presentation of the $\mu$-calculus and some of its
fragments is taken from \cite{HM00}.

\begin{definition}
The formulas of the {\em $\mu$-calculus} over a set of
observables $P$ are generated by the grammar
$$
\varphi ::= p \mid \overline{p} \mid h \mid \varphi \vee \varphi \mid \varphi
\wedge \varphi \mid \existsnext \varphi \mid \forallnext \varphi \mid
(\mu h : \varphi) \mid (\nu h : \varphi)
$$
for $p \in P$ and variables $h$ from some fixed set.
\end{definition}

For functions $\Epsilon$, we write $\Epsilon [ h \mapsto \tau ]$ for
the mapping that agrees on $\Epsilon$ on all values in its domain,
except that $h$ is instead mapped to $\tau$. Given a transition system
${\cal S} = (Q, \delta, \ext{\cdot}, P)$, and a mapping from the
variables to sets of states $\Epsilon$, any formula $\varphi$ of
the $\mu$-calculus over $P$ defines a set $\syns{\varphi}{\Epsilon}
\subseteq Q$ of states:
$$
\begin{array}{l}
\syns{p}{\Epsilon} = \ext{p} \\
\syns{\overline{p}}{\Epsilon} = Q \setminus \ext{p} \\
\syns{h}{\Epsilon} = \Epsilon(h) \\
\syns{\varphi_1 \switch{\vee}{\wedge} \varphi_2}{\Epsilon} =
\syns{\varphi_1}{\Epsilon} \switch{\cup}{\cap} \syns{\varphi_2}{\Epsilon} \\
\syns{\switch{\exists}{\forall} \bigcirc \varphi}{\Epsilon} = \{ s \in Q
\mid \switch{\exists}{\forall} s' \in \delta(s) : s' \in
\syns{\varphi}{\Epsilon} \} \\
\syns{\switch{\mu}{\nu} h : \varphi}{\Epsilon} = \switch{\cap}{\cup} \{
\tau \subseteq Q \mid \tau = \syns{\varphi}{\Epsilon[h \mapsto \tau]} \}.
\end{array}
$$

The logic $L_1^\mu$ over a set of observables $P$ is the set of closed
formulas of the $\mu$-calculus over $P$.  We will write ${\cal S}, s
\models \varphi$ when $s \in \syns{\varphi}{\Epsilon}$ for any
$\Epsilon$. (As an $L^\mu_1$ formula $\varphi$ is closed, the initial
mappings in $\Epsilon$ are never used and the validity is therefore
independent of $\Epsilon$.)

Usually we are not interested in which states satisfy a given formula,
rather we want to know whether the set of initial states of a system
satisfy it or not. We therefore introduce a notion of satisfaction and
write ${\cal S}, b_0 \models \varphi$, where $b_0$ is a boolean
variable of $\cal P$, to mean that for all states $s \in \ext{b_0}$,
we have ${\cal S}, s \models \varphi$.

We will also use the following fragments of the $\mu$-calculus:

\begin{definition}
  The logic $L_2^\mu$ (the {\em existential fragment of the
    $\mu$-calculus}) is the subset of $L_1^\mu$ without the
  constructors $\overline{p}$ or $\forallnext$.
\end{definition}

\begin{definition}
The logic $L_4^\mu$ (the {\em existential conjunction-free fragment of
  the $\mu$-calculus}) is the subset of $L_2^\mu$ without the
constructors $\wedge$ or $\nu$. 
\end{definition} 

$L^\mu_1$ is strictly more expressive than $L^\mu_2$, which is
strictly more expressive than $L^\mu_4$ \cite{HM00}.\footnote{The
  logics $L^\mu_3$ (linear-time $\mu$-calculus) and $L^\mu_5$
  (reachability) are not required in this paper.}

For any logic $L^\mu_i$, there is a {\em dual logic} $\overline{L^\mu_i}$
obtained by replacing the constructors $p, \overline{p}, \vee, \wedge,
\existsnext, \forallnext, \mu, \nu$ in formulas $\varphi$ by
$\overline{p}, p, \wedge, \vee, \forallnext,\existsnext,
\nu, \mu$ respectively to form formulas $\overline{\varphi}$. The
satisfaction of an $\overline{L^\mu_i}$ formula $\overline{\varphi}$
by a state $s \in Q$ is complementary to the satisfaction of the
formula $\varphi$ in the logic $L^\mu_i$ by $s$, ie.\ ${\cal S}, s
\models \varphi$ iff ${\cal S}, s \not \models \overline{\varphi}$.

\section{Language of programs}

\label{sect:lang}

Here we define the syntax of our programs, which is based on that of
UNITY \cite{CM88}. It is a language of guarded multiple assignments,
extended with simple array operations. We give semantics to these
programs in terms of transition systems.

Our programs are data-independent with respect to a set of type
symbols, as the only operations they allow on values of these types
are non-deterministic selection (with no assumption of fairness),
copying between variables, and equality testing. In addition, they may
read and write these values to arrays indexed by other such type
symbols.

We then describe the subclass of these programs we will be considering
in this paper and the problem we will be addressing.

\subsection{Syntax}

We assume the existance of a set of symbols called {\em type symbols}.

A {\em program} $\cal P$ is:
\begin{itemize}
\item A finite set of {\em variables} together with their {\em types},
  partitioned into three sets:
  \begin{itemize} \item boolean variables, of type $\mathbbm{B}$,
  \item data variables, of type $Z$ where $Z$ is some type symbol,
  \item array variables, of type $\Array{X}{Y}$ where $X$ and $Y$ are
    type symbols.
  \end{itemize}
\item A finite set of guarded commands $e \longrightarrow I$,
  where:
  \begin{itemize}
  \item The {\em boolean expression} $e$ is taken from the grammar
    $$
    e ::= \true \mid \false \mid b \mid z = z' \mid \neg e \mid e \vee e,
    $$
    where $b$ ranges over the boolean variables, and $z$ and $z'$ are
    data variables of the same type.
  \item The {\em command} $I$, representing a simultaneous multiple
    assignment, is a set containing at most:
    \begin{itemize}
    \item for each boolean variable $b$, an assignment $b := e$, where
      $e$ is a boolean expression,
    \item for each data variable $z$ of type $Z$, at most one of $z :=
      z'$, $z :=\ ?$, or $\Read{z}{a}{x}$, where $z'$, $a$, and $x$
      are any variables with types $Z$, $\Array{X}{Z}$, and $X$
      respectively for some type symbol $X$,
    \item for each array $a$ of type $\Array{X}{Y}$, an operation
      $\Write{a}{x}{y}$, where $x$ and $y$ are variables of type $X$
      and $Y$ respectively.
    \end{itemize}
  \end{itemize}
\end{itemize}

{\bf Notation:} We may write multiple assignments as two lists of
equal length separated by $:=$, eg.\ $x, y := y, x$ repesents the
multiple assignment consisting of both $x := y$ and $y := x$. We may
also denote the array operations $\Read{y}{a}{x}$ and
$\Write{a}{x}{y}$ with the C-like syntaxes $y := a[x]$ and $a[x] := y$
respectively.

\subsection{Semantics}

A {\em type instantiation} ${\cal I}$ for a program $\cal P$ is a
function from the type symbols in $\cal P$ to non-empty sets upon
which equality is decidable.

The semantics of a program $\cal P$ together with a type
instantiation ${\cal I}$ for it, denoted $\semi{\cal P}{\cal I}$, is the
transition system $(Q, \delta, \ext{\cdot}, P)$, where:
\begin{itemize}
\item The state space $Q$ is the set of all total functions from the
  variables of $\cal P$ into
  \begin{itemize}
  \item for boolean variables, the set $\mathbbm{B} = \{ \true, \false \}$,
  \item for data variables of type $Z$, the set ${\cal I}(Z)$,
  \item for array variables of type $\Array{X}{Y}$, the total-functions
    space ${\cal I}(X) \rightarrow {\cal I}(Y)$.
  \end{itemize}
\item $s' \in \delta(s)$ if and only if there is some guarded command
  $e \longrightarrow I$ in $\cal P$ such that $E_s(e) = \true$ and $s
  \Delta_I s'$ where:
  \begin{itemize}
  \item The evaluating function $E$ for a boolean expression in a state $s$ is
    defined as follows:
    $$
    \begin{array}{lcl}
    E_s(\true) &=& \true, \\
    E_s(\false) &=& \false, \\
    E_s(e_1 \vee e_2) &=& E_s(e_1) \mbox{ `or' } E_s(e_2), \\
    E_s(\neg e_1) &=& \mbox{`not' } E_s(e_1), \\
    E_s(b) &=& s(b), \\
    E_s(z = z') &=& ( s(z) = s(z') ),
    \end{array}
    $$
    for boolean variables $b$ and data variables $z$ and $z'$.
  \item The relation $\Delta_I$ on pairs of states for a multiple
    assignment $I$ is defined as $s \Delta_I s'$ if and only if all
    of the following:
    \begin{itemize}
    \item for each boolean variable $b$, if $b := e$ is in $I$, then
      $s'(b) = E_s(e)$, else $s'(b) = s(b)$,
    \item for each data variable $z$, \\
      \begin{tabular}{l}
      if $z := z'$ is in $I$, then $s'(z) = s(z')$, \\
      else if $\Read{z}{a}{x}$ is in $I$, then $s'(z) = s(a)(s(x))$, \\
      else either $z :=\ ?$ is in $I$ or $s'(z) = s(z)$,
      \end{tabular}
    \item for each array variable $a$ of type $\Array{X}{Y}$, and for
      each $v \in {\cal I}(X)$,\\
    \begin{tabular}{l}
    if there are $x$ and $y$ variables such that \\ \quad
      $\Write{a}{x}{y}$ is in $I$ and $s(x) = v$, then $s'(a)(v) = s(y)$, \\
      else $s'(a)(v) = s(a)(v)$.
    \end{tabular}
    \end{itemize}
  \end{itemize}
\item the observables $P$ is the set of boolean variables,
\item the extensions function is defined as
  $$
  \ext{b} = \{ s \in Q \mid s(b) = \true \}.
  $$
\end{itemize}

{\bf Notation:} We may write $s(a[x])$ to mean $s(a)(s(x))$ for
states $s$, array variables $a$, and data variables $x$.

It can be noticed that it is only the {\em cardinalities} of the type
instances which affect the observable semantics. Formally, given two
type instantiations ${\cal I}_1$ and ${\cal I}_2$ for a program $\cal
P$, where $|{\cal I}_1(Z)| = |{\cal I}_2(Z)|$ for all type symbols $Z$
in $\cal P$, there exists a bisimulation $\rightleftharpoons$ between
${\cal S}_1 = \semi{\cal P}{{\cal I}_1}$ and ${\cal S}_2 = \semi{\cal
  P}{{\cal I}_2}$.  This is because the observable semantics depend on
the equality relationships on values of these types, and bijections
preserve equality. If ${\cal I}_1(Z)$ and ${\cal I}_2(Z)$ have the
same cardinality, then there exists a bijection $f_Z$ between them;
the bisimulation $\rightleftharpoons$ uses these bijections to
translate values between ${\cal S}_1$ and ${\cal S}_2$.

It follows that, for the results in this paper, any type instantiation
$\cal I$ can be replace by $\cal I'$ which maps onto an initial
portion of the cardinal numbers, ie.\ ${\cal I}'(Z) = \{1, \ldots,
|{\cal I}(Z)| \}$.

\subsection{This paper}

For simplicity, in this paper we consider programs with only two type
symbols $X$ and $Y$, and array variables only of type $\Array{X}{Y}$.
We will write $\semi{\cal P}{A, B}$ as shorthand for $\semi{\cal
  P}{\cal I}$ where $\cal I$ maps $X$ and $Y$ to the sets $A$ and $B$
respectively.

In particular we do not consider the extension of this language to
include the array reset operation, which assigns a given value of type
$Y$ to all array components. The operational semantics of such an
operation would dictate that the successor state maps the array
variable to the constant function returning the $Y$ value. Array reset
is too expressive to obtain results as powerful as we do here
\cite{RL01}.

We will use variables $b$, $b'$, $b_i$, ... to denote variables of
type $\mathbbm{B}$, and similarly $x$, $y$ and $a$ for variables of
type $X$, $Y$ and $\Array{X}{Y}$ respectively. We will also use $z$
for variables of either type $X$ or $Y$, and $e$ for boolean
expressions.

The main problem of interest is the following {\em parameterised
  model-checking problem}: given a data-independent program $\cal P$
with arrays , a boolean variable $b_0$ of $\cal P$, and a temporal
logic formula $\varphi$ referring to control states of $\cal P$, is it
true that $\semi{\cal P}{\cal I}, b_0 \models \varphi$ for all type
instantiations $\cal I$ which map $X$ and $Y$ to non-empty finite
sets.

\begin{example}
\label{ex:singleCache}
Our example programs will use variables that range over finite
datatypes, such as program counters, even though these are not part of
our formally considered language.  This is because such values can be
coded as tuples of booleans, which are allowed.  Similarly we will use
familiar programming constructs such as if-then-else, goto, and
nondeterministic choice \verb$|~|$ because the effects of these can be
achieved using guarded commands and booleans.

Figure \ref{fig:singleCache1} shows a fault-tolerant interface over a
set of unreliable memories, which we expect to work provided there is
no more than one error. It is parameterised by two types \code{ADDR}
and \code{DATA} representing the types of addresses and data values
respectively, and the program is data independent with arrays without
reset with respect to these types.  The memories are represented by
arrays called \code{mem1}, \code{mem2} and \code{mem3}, and the
address and data busses are represented by the variables
\code{addrBus} and \code{dataBus}.

In \code{LOOP}, values appear on the address and data busses and are
used to write to or read from memory. When writing to memory, the data
value is written to all three arrays at the appropriate place. When
reading from memory, the program takes the majority value of all three
memories at that location if such a value exists.

We have incorporated the faulty behaviour of the memories into our
program. Of course this would not be present in the final code, but
our arrays are not naturally faulty so we need to simulate that
behaviour in order to do any interesting analysis on our program. So,
in between reads and writes, a fault may occur which writes a
nondeterministic value to one of the memories at any location.

A property we would usually desire of a memory system is that a read
from an arbitrary location will always return the value of the last
write to that location, provided there has been one. Because of the
possibility of faults in this system, we would expect this to be true
until two faults have occurred.

Figure \ref{fig:singleCache2} shows the code again, annotated with
``checking code'' marked with \verb$#$'s. This code unobtrusively
monitors the progress of the system and moves it to a special
\code{ERROR} state when it detects that the program's specification
has been broken. The new code requires its own variables:
\code{testAddr} holds the arbitrary memory location which is being
monitored and \code{testData} contains the last value written there,
provided that \code{testWritten} is true. The variable \code{faults}
records whether the number of faults so far is none, one, or more than
one. The annotations in the code maintain these invariants.

\begin{figure}
{\footnotesize
\begin{verbatim}
VARIABLES:
    addrBus: ADDR
    dataBus: DATA
    data1, data2, data3: DATA
    mem1, mem2, mem3: DATA[ADDR]

START:
    goto LOOP

LOOP:
    addrBus, dataBus := ?, ?
    goto READ |~| goto WRITE |~| goto FAULT

READ:
    data1, data2, data3 := mem1[addrBus], mem2[addrBus], mem3[addrBus]
    if data1 != data2 then dataBus := data3 else dataBus := data1
    goto LOOP

WRITE:
    mem1[addrBus], mem2[addrBus], mem3[addrBus] := dataBus, dataBus, dataBus
    goto LOOP

FAULT:
    mem1[addrBus] := dataBus |~| mem2[addrBus] := dataBus
            |~| mem3[addrBus] := dataBus
    goto LOOP
\end{verbatim}}
\caption{Fault-tolerant memory.
\label{fig:singleCache1}}
\end{figure}

\begin{figure}
{\footnotesize
\begin{verbatim}
VARIABLES:
1   addrBus: ADDR
2   dataBus: DATA
3   data1, data2, data3: DATA
4   mem1, mem2, mem3: DATA[ADDR]
5#  testAddr: ADDR
6#  testData: DATA
7#  testWritten: BOOL
8#  faults: {0..2}

START:
1#  faults, testWritten := 0, false
2   goto LOOP

LOOP:
1   addrBus, dataBus := ?, ?
2   goto READ |~| goto WRITE |~| goto FAULT

READ:
1   data1, data2, data3 := mem1[addrBus], mem2[addrBus], mem3[addrBus]
2   if data1 != data2 then dataBus := data3 else dataBus := data1
3#  if addrBus = testAddr and testWritten and faults < 2 and dataBus != testData
            then goto ERROR
4   goto LOOP

WRITE:
1   mem1[addrBus], mem2[addrBus], mem3[addrBus] := dataBus, dataBus, dataBus
2#  if addrBus = testAddr then testData, testWritten := dataBus, true
3   goto LOOP

FAULT:
1   mem1[addrBus] := dataBus |~| mem2[addrBus] := dataBus
            |~| mem3[addrBus] := dataBus
2#  if faults < 2 then faults := faults + 1
3   goto LOOP

ERROR:
1#  goto ERROR
\end{verbatim}}
\caption{Fault-tolerant memory composed with specification.
\label{fig:singleCache2}}
\end{figure}

In order to test that the system satisfies its specification, we need
to check that the \code{ERROR} state is never reachable from the start,
whatever finite non-empty sets $A$ and $B$ are used as instances of
\code{ADDR} and \code{DATA}. This can be expressed using
$\overline{L^\mu_4}$ as
$$\forall A, B \cdot \semi{\cal P}{A, B}, b_0 \models \nu h:
\forallnext (\overline{b_E} \wedge h),
$$ where $b_0$ is a special boolean variable of the program that must be
true for the program line \code{START} to be executed, where it is set
to false, and must be false for all other guarded instructions;
similarly, $b_E$ must be false for all instructions, and is set to
true at the line \code{ERROR}.
\end{example}

\section{Infinite arrays}

\label{sect:infinite}

In this section we consider the class of systems where $X$ and $Y$ are 
both instantiated to infinite sets.

We provide a syntactic translation from programs with arrays to
programs without arrays. We show that there exists a bisimulation
between the former with semantics that use partial functions with
finite domains to model arrays, and the latter with normal semantics.
From this, we deduce that the $\mu$-calculus model-checking problem is
decidable for this class of systems.

This section is organised into the following subsections.  The
partial-functions semantics is introduced in \ref{sub:partial}; the
translation is described in \ref{sub:translation}; the bisimulation
and its proof are in \ref{sub:connection}; the model-checking result
is deduced in \ref{sub:conclusion}.

\subsection{Partial-functions semantics}

\label{sub:partial}

For infinite instantiations for $X$ and $Y$, the semantic values for
arrays are finite partial functions.  An undefined location in an
array represent nondeterminism which is yet to be resolved; this
nondeterminism is resolved exactly when the system inputs the
corresponding index value into one of its variables. These semantics
are formalised here.

The {\em partial-functions semantics} of a program $\cal P$ together
with a type instantiation ${\cal I}$ for it, denoted $\pfsem{\cal
P}{\cal I}$, is the transition system $(Q^*, \delta^*,
\ext{\cdot}^*, P)$, which differs from the normal semantics as follows:
\begin{itemize}
\item A state $s \in Q^*$ maps array variables to {\em finite partial}
functions (ie.\ defined only on a finite subset of their domains)
instead of total functions, but we insist that, for all array
variables $a$ with type $\Array{X}{Y}$, the partial function $s(a)$ is
defined at $s(x)$ for all variables $x$ of type $X$.
\item The relation $\Delta_I$ is amended to $\Delta^*_I$ so that
  $s \Delta^*_I s'$ imposes a different condition for array variables:
  \begin{itemize}
  \item for each array variable $a$, and for each $v \in {\cal I}(X)$,\\
    \begin{tabular}{l}
    if there are variable $x$ and $y$ such that \\ \quad
      $\Write{a}{x}{y}$ is in $I$ and $s(x) = v$, then $s'(a)(v) = s(y)$, \\
      else if there does not exist an $x$ variables such that \\
      \quad $x :=\ ?$ is in $I$ and $s'(x) = v$ and $s(a)(v) = \bot$, \\
      \quad then $s'(a)(v) = s(a)(v)$.
    \end{tabular}
  \end{itemize}
\end{itemize}
(Note that the final ``if'' has no ``else'' case --- ie.\ the
statement holds when the ``if'' condition is false.)  The ``else''
clause of the arrays case above could be read as follows: if there is
a variable $x$ which is non-deterministically selected to $v$ during
the transition, where $a$ was undefined at $v$ before, then the new
value of $a$ at $v$ is unspecified; otherwise it must remain the same.

{\bf Notation:} We write $f(v) = \bot$ to mean $f$ is undefined at
$v$, and use the conventions that $\bot = \bot$ and $\bot \neq w$ for
any value $w$.

\subsection{Equivalent programs without arrays}

\label{sub:translation}

Here we provide a syntactic translation from programs with arrays to
programs without arrays.

We begin by extending our language slightly to allow sequences of
guarded commands to be executed in one atomic transition. Note we say
{\em command} to mean the multiple assignment $I$ in a guarded command
$e \longrightarrow I$.

\begin{definition}
We can {\em append} a guarded command $e_2 \longrightarrow I_2$ onto a
command $I_1$, to form a single command $I_1 : e \longrightarrow I_2$.
The semantics of the new command are $s \Delta_{I_1 : e
  \longrightarrow I_2} s''$ if and only if either
\begin{itemize}
\item there exists $s'$ such that $s \Delta_{I_1} s'$ and $s'
  \Delta_{I_2} s''$ and $E_{s'}(e) = \true$, or 
\item $s \Delta_{I_1} s''$ and $E_{s''}(e) = \false$.
\end{itemize}
\end{definition}
Note it is possible to append many guarded commands onto a single
command.

We will also need to split commands into two as follows: a command $I$
can be split into its $X$-type assignments $I_X$ and $Y$-type and
boolean assignments $I_Y$ as follows:
\begin{itemize}
\item $I_X$ contains exactly all the assignments of the form $x :=
  x'$ and $x :=\ ?$ from $I$.
\item $I_Y$ contains exactly all the assignments of the form $y :=
  y'$, $y :=\ ?$, $\Write{a}{x}{y}$, $\Read{y}{a}{x}$, $b := e$ from $I$.
\end{itemize}

We now provide the syntactic translation from programs with arrays to
programs without arrays. From a program ${\cal P}$, we can form its
{\em array-free abstraction} $\cal P^\sharp$ as follows.
\begin{itemize}
\item For each array $a$ and each variable $x$ of type $X$, we add a
  new variable of type $Y$, which we will call $ax$.
\item $\cal P^\sharp$ contains no arrays.
\item Perform a translation on each command $I$ to form a
  new command $I^\sharp_Y : \true \longrightarrow I^\sharp_X$ as follows:
  \begin{enumerate}
  \item The multiple assignment $I^\sharp_Y$ is the same as $I_Y$ except:
    \begin{itemize}
    \item for each $\Read{y}{a}{x}$ appearing in $I$, we instead
      have $y := ax$ in $I^\sharp_Y$;
    \item for each $\Write{a}{x}{y}$ appearing in $I$, we instead
      have $ax := y$ in $I^\sharp_Y$.
    \end{itemize}
    For each $\Write{a}{x}{y}$ appearing in $I$, append onto
    $I^\sharp_Y$ the following guarded command for each other variable
    $x'$ (in any order):
    $$
    x = x' \longrightarrow ax' := ax.
    $$
  \item The multiple assignment $I^\sharp_X$ is the same as $I_X$ except:
    \begin{itemize}
    \item for each $x := x'$ appearing in $I$, we also have $ax :=
      ax'$ in $I^\sharp_X$ for all arrays $a$.
    \item for each $x :=\ ?$ appearing in $I$, we also have $ax :=\ ?$
      in $I^\sharp_X$ for all arrays $a$.
    \end{itemize}
    For each $x :=\ ?$ appearing in $I$, append onto $I^\sharp_X$ the
    following guarded command for each other variable $x'$ of type $X$
    such that $x' :=\ ?$ is not in $I$ (in any order):
    $$
    x = x' \longrightarrow a_1 x, ..., a_l x := a_1 x', ..., a_l x'
    $$
    for all the arrays $a_1, ..., a_l$.
    
    Let $x_1, \ldots, x_n$ be any enumeration of all the variables of
    type $X$ such that $x :=\ ?$ appears in $I$. Append further onto
    $I^\sharp_X$, for each pair $i$ and $j$ both from $1$ to $n$ such
    that $i > j$, in lexicographical order of $(i, j)$, the guarded
    command:
    $$
    x_i = x_j \longrightarrow a_1 x_i, ..., a_l x_i := a_1 x_j, ..., a_l x_j
    $$
    for all the arrays $a_1, ..., a_l$.
  \end{enumerate}
\end{itemize}


\begin{figure}
{\footnotesize
\begin{verbatim}
VARIABLES:
1   addrBus: ADDR
2   dataBus: DATA
3   data1, data2, data3: DATA
4   mem1_addrBus, mem1_testAddr, mem2_addrBus, mem2_testAddr, mem3_addrBus,
            mem3_testAddr: DATA
5#  testAddr: ADDR
6#  testData: DATA
7#  testWritten: BOOL
8#  faults: {0..2}

START:
1#  faults, testWritten := 0, false
2   goto LOOP

LOOP:
1   dataBus := ?
        : addrBus, mem1_addrBus, mem2_addrBus, mem3_addrBus := ?, ?, ?, ?
        : if addrBus = testAddr then mem1_addrBus, mem2_addrBus, mem3_addrBus :=
            mem1_testAddr, mem2_testAddr, mem3_testAddr
2   goto READ |~| goto WRITE |~| goto FAULT

READ:
1   data1, data2, data3 := mem1_addrBus, mem2_addrBus, mem3_addrBus
2   if data1 != data2 then dataBus := data3 else dataBus := data1
3#  if addrBus = testAddr and testWritten and faults < 2 and dataBus != testData
            then goto ERROR
4   goto LOOP

WRITE:
1   mem1_addrBus, mem2_addrBus, mem3_addrBus := dataBus, dataBus, dataBus
        : if addrBus = testAddr then mem1_testAddr := mem1_addrBus
        : if addrBus = testAddr then mem2_testAddr := mem2_addrBus
        : if addrBus = testAddr then mem3_testAddr := mem3_addrBus
2#  if addrBus = testAddr then testData, testWritten := dataBus, true
3   goto LOOP

FAULT:
1   mem1_addrBus := dataBus
        : if addrBus = testAddr then mem1_testAddr := mem1_addrBus
    |~|
    mem2_addrBus := dataBus
        : if addrBus = testAddr then mem2_testAddr := mem2_addrBus
    |~|
    mem3_addrBus := dataBus
        : if addrBus = testAddr then mem3_testAddr := mem3_addrBus
2#  if faults < 2 then faults := faults + 1
3   goto LOOP

ERROR:
1#  goto ERROR
\end{verbatim}}
\caption{Array-free abstraction of fault-tolerant memory composed with
specification.
\label{fig:singleCacheTransl}}
\end{figure}

\begin{example}
\label{ex:singleCacheTransl}
The array-free abstraction of Example \ref{ex:singleCache} is shown in
Figure \ref{fig:singleCacheTransl}. Note the use of the append
operator \verb$:$ to group together instructions into one atomic
transition.
\end{example}

\subsection{The connection}

\label{sub:connection}

We now identify the relationship between a program $\cal P$ and its
array-free abstraction $\cal P^\sharp$.  We show that, for infinite
instantiations for the types $X$ and $Y$, there exists a bisimulation
between the transition system produced using partial-functions
semantics on $\cal P$ and the transition system produced using normal
semantics on $\cal P^\sharp$. We first present some auxiliary
definitions.

\begin{definition}
The set $\TERMS_X$ is the set of variables of type $X$, and if we
write $s(\TERMS_X)$, it means the set $\{ s(x) \mid x \in \TERMS_X
\}$.  An {\em $X$-bijection} $\alpha$ on two states $s$ and $t$ is a
bijection $\alpha : s(\TERMS_X) \rightarrow t(\TERMS_X)$ such that
$\alpha(s(x)) = t(x)$ for all variables $x$ of type $X$.
\end{definition}

Given a program ${\cal P}$ and two infinite sets $A^*$ and
$B^*$, let
$$
\begin{array}{rlcl}
& \pfsem{\cal P}{A^*, B^*} &=& (Q^*, \delta^*,
\ext{\cdot}^*, P) \\
\mbox{ and } & \semi{\cal P^\sharp}{A^*, B^*} &=&
  (Q, \delta, \ext{\cdot}, P).
\end{array}
$$

\begin{definition}
\label{def:approx}
We define the relation $\approx \: \subseteq Q \times Q^*$ as $s \approx
t$ exactly when
\begin{itemize}
\item $s(b) = t(b)$ for boolean variables $b$,
\item there exists a $X$-bijection on $s$ and $t$,
\item $s(y) = t(y)$, for all variables $y$ of type $Y$, and
\item $s(ax) = t(a[x])$, for all arrays $a$ and $X$-variables $x$.
\end{itemize}
Note that the range of $\approx$ is the whole of $Q^*$, while the
domain of $\approx$ is only the states $s$ in $Q$ that satisfy the
{\em array-consistency formula}
$$
\Sigma \quad \equiv \quad \forall x, x' \cdot x = x' \implies
\forall a \cdot ax = ax'.
$$
\end{definition}

Our aim is to prove that $\approx$ is a bisimulation. The proof relies
on the following observation about $\pfsem{\cal P}{A^*, B^*}$: when a
value $v$ of type $X$ is forgotten by the program (ie.\ it is
overwritten in one of the variables of type $X$), the program's
behaviour is unaffected if it never sees $v$ again, and so the
corresponding $Y$-values in the arrays may also be forgotten. It
therefore only needs to remember the parts of the array currently in
view --- a finite number of values.

This may appear to cause problems, because in reality that value could
later be reintroduced (using $x :=\ ?$), and values from the arrays at
$v$ then read.  For an accurate model, these values would have to
equal those originally written into the array, which the abstraction
$\cal P^\sharp$ has forgotten.  However, as the arrays are always
undefined at places, an indistinguishable behaviour could happen
anyway if a brand new $X$-value was chosen and the non-determinism was
resolved in an appropriate way.  Because the program is
data-independent with respect to $X$, it has no way of telling that
the new value is not the forgotten $v$.

It is the $X$-bijection in the relation above that allows us to switch
this forgotten value for a brand new one. The data independence of $Y$
is not actually required here, but is used later to model check $\cal
P^\sharp$.

First, we present a result which allows us to break a command up into
more manageable pieces.
  
\begin{lemma}
  \label{lem:comp}
  For $s_1, s_3 \in Q^*$, we have $s_1 \Delta^*_I s_3$ if and only if
  there exists a state $s_2 \in Q^*$ such that $s_1 \Delta^*_{I_Y}
  s_2$ and $s_2 \Delta^*_{I_X} s_3$.
\end{lemma}
\begin{proof*} $\Rightarrow$:
  Define $s_2$ as follows:
  $$
  \begin{array}{rcl}
    s_2(b) &=& s_3(b), \mbox{ for $b$ of type $\mathbbm B$},\\
    s_2(x) &=& s_1(x), \mbox{ for $x$ of type $X$},\\
    s_2(y) &=& s_3(y), \mbox{ for $y$ of type $Y$},\\
    s_2(a)(v) &=& s_1(y), \mbox{ if $\Write{a}{x}{y}$ is in $I$ and
      $s_1(x) = v$},\\
    &=& s_1(a)(v), \mbox{ otherwise}.
  \end{array}
  $$
  Now we prove that $s_1 \Delta^*_{I_Y} s_2$:
  \begin{itemize}
  \item $b := e$ in $I_Y$ implies it's also in $I$, so $s_2(b) = s_3(b)
    = s_1(e)$; else $s_2(b) = s_3(b) = s_1(b)$.
  \item There are no $x := x'$ or $x :=\ ?$ in $I$, and $s_2(x) =
    s_1(x)$ by definition.
  \item If $y := y'$ in $I_Y$ then it's also in $I$, so $s_2(y) = s_3(y) =
    s_1(y)$; else $\Read{y}{a}{x}$ in $I_Y$ implies it's also in $I$, so
    $s_2(y) = s_3(y) = s_1(a)(s_1(x))$; else, if $y :=\ ?$ is not in $I_Y$
    then it's not in $I$ either, so $s_2(y) = s_3(y) = s_1(y)$.
  \item For each array $a$ and $v \in A^*$
    \begin{itemize}
    \item If $\Write{a}{x}{y}$ is in $I$ and $s_1(x) = v$, then
      $s_2(a)(v) = s_1(y)$ by definition.
    \item Else $s_2(a)(v) = s_1(a)(v)$ as there is no $x :=\ ?$ in
      $I_Y$.
    \end{itemize}
  \end{itemize}
  Now we prove that $s_2 \Delta^*_{I_X} s_3$:
  \begin{itemize}
  \item There is no $b := e$ in $I_X$, and $s_3(b) = s_2(b)$ by
    definition of $s_2$.
  \item If $x := x'$ in $I_X$ then it's in $I$, and so $s_3(x) =
    s_1(x') = s_2(x')$; else if $x :=\ ?$ is not in $I_X$, then it's
    not in $I$, so $s_3(x) = s_1(x) = s_2(x)$.
  \item For each array $a$ and $v \in A^*$
    \begin{itemize}
    \item There is no $\Write{a}{x}{y}$ in $I_X$.
    \item so assume there does not exist an $x :=\ ?$ in $I_X$ such
      that $s_3(x) = v$ and $s_2(a)(v) = \bot$. Then there does not
      exist such an $x :=\ ?$ in $I$ such that $s_3(x) = v$ and
      $s_1(a)(v) = s_2(a)(v) = \bot$ (by definition of $s_2$), so
      $s_3(a)(v) = s_1(a)(v) = s_2(a)(v)$ (again by definition of
      $s_2$).
    \end{itemize}
  \end{itemize}
  
  $\Leftarrow$: Assume $s_1 \Delta^*_{I_Y} s_2$ and $s_2 \Delta^*_{I_X}
  s_3$. We will now prove $s_1 \Delta^*_I s_3$:
  \begin{itemize}
  \item If $b := e$ is in $I$ then $b := e$ is in $I_Y$, so $s_2(b) =
    E_{s_1}(e)$. There are no boolean assignments in $I_X$ so $s_3(b)
    = s_2(b)$; if $b := e$ is not in $I$, then there are no boolean
    assignments in either $I_Y$ or $I_X$, so $s_3(b) = s_1(b)$.
  \item The cases for data variables are very similar to those for
    boolean variables.
  \item For each array variable $a$ and each $v \in A^*$,
    \begin{itemize}
    \item If there are $x$ and $y$ variables such that
      $\Write{a}{x}{y}$ is in $I$ and $s_1(x) = v$, then
      $\Write{a}{x}{y}$ will also appear in $I_Y$.  There are no
      writes in $I_X$, and $s_2(a)(s_2(x))$ can not be $\bot$ so
      $s_3(a)(v) = s_2(a)(v)$.  We get $s_3(a)(v) = s_2(a)(v) = y$.
    \item Otherwise assume there does not exist an $x :=\ ?$ in $I$
      such that $s_3(x) = v$ and $s_1(a)(v) = \bot$. Then there cannot
      exists an $x :=\ ?$ in $I_X$ such that $s_3(x) = v$ and
      $s_2(a)(v) = \bot$, because $s_2(a)(v) = s_1(a)(v)$ (no
      $\Write{a}{x}{y}$ in $I_Y$). Therefore, we have $s_3(a)(v) =
      s_2(a)(v) = s_1(a)(v)$. \mathproofbox
    \end{itemize}
  \end{itemize}
\end{proof*}

In the following five lemmas, which all have $s \approx t$ as a premise,
let $\alpha$ be the $X$-bijection from $s$ to $t$.

\begin{lemma}
  \label{lem:E}
  If $s \approx t$, then $E_s(e) = E_t(e)$ for any boolean expression $e$.
\end{lemma}
\begin{proof}
  From $s \approx t$, we know
  \begin{itemize}
  \item $E_s(b) = E_t(b)$, because $s(b) = t(b)$ for all booleans
    variables $b$,
  \item $E_s(y = y') = E_t(y = y')$, because $s(y) = t(y)$ for all
    variables $y$ of type $Y$,
  \item and also,
    $$
    \begin{array}{rl}
      & E_t(x = x')\\
      =& (t(x) = t(x'))\\
      =& (\alpha(s(x)) = \alpha(s(x')))\\
      =& \pstep{$\alpha$ is a bijection}\\
      & (s(x) = s(x'))\\
      =& E_s(x = x').
    \end{array}
    $$
  \end{itemize}
  By structural induction on $e$, using the above as base cases, it
  can easily be shown that $E_s(e) = E_t(e)$.
\end{proof}

\begin{lemma}
  \label{lem:Y1}
  If $s \approx t$ and $s \Delta_{I^\sharp_Y} s'$, then there exists $t'$
  such that $s' \approx t'$ and $t \Delta^*_{I_Y} t'$.
\end{lemma}
\begin{proof*}
  Define $t'$ as follows:
  $$
  \begin{array}{rcl}
    t'(b) &=& s'(b),\\
    t'(x) &=& \alpha(s'(x)),\\
    t'(y) &=& s'(y),\\
    t'(a)(v) &=& s'(ax), \mbox{ if there is such an $x$ where $t'(x)
      = v$},\\
    &=& t(y), \mbox{ (else) if $\Write{a}{x}{y}$ is in $I_Y$ and
      $t(x) = v$},\\
    &=& t(a)(v), \mbox{ otherwise}.\\
  \end{array}
  $$
  We need to show that the first case for arrays is well-defined, that 
  is: if $t'(x) = t'(x')$, then $s'(ax) = s'(ax')$. First notice:
  $$
  \begin{array}{rrcl}
    & t'(x) &=& t'(x')\\
    \implies & \alpha(s'(x)) &=& \alpha(s'(x'))\\
    \implies & && \pstep{$\alpha$ is a bijection}\\
    & s'(x) &=& s'(x').
  \end{array}
  $$
  Assuming $s'(x) = s'(x')$, it can be seen that if there is some
  $y$ such that $\Write{a}{x}{y}$ or $\Write{a}{x'}{y}$ are in $I_Y$,
  then the appendages on $I^\sharp_Y$ will make sure that $s'(ax) =
  s'(ax')$.  If there are no writes to $a[x]$ nor $a[x']$ then both
  $ax$ and $ax'$ are unaffected between $s$ and $s'$, and we get
  $$
  \begin{array}{rrcl}
    & t'(x) &=& t'(x')\\
    \implies & && \pstep{$x$ and $x'$ not affected in $I_Y$}\\
    & t(x) &=& t(x')\\
    \implies & t(a)(t(x)) &=& t(a)(t(x'))\\
    \implies & t(a[x]) &=& t(a[x'])\\
    \implies & && \pstep{$s \approx t$}\\
    & s(ax) &=& s(ax')\\
    \implies & && \pstep{$ax$ and $ax'$ unaffected in $I^\sharp_Y$}\\
    & s'(ax) &=& s'(ax')
  \end{array}
  $$
  
  From the definition of $t'$, notice that $t'(a[x]) = s'(ax)$ for all
  $x$.  Notice further that $s' \approx t'$.
  
  We now wish to show that $t \Delta^*_{I_Y} t'$. We will run through the
  cases from the definition of $\Delta^*$.
  \begin{itemize}
  \item For any boolean variables $b$, either we have (a) $b := e$ in
    $I_Y$, in which case $b := e$ also appears in $I^\sharp_Y$ as the
    only assignment to $b$, so $t'(b) = s'(b) = E_{s}(e) = E_{t}(e)$
    (the last step by Lemma \ref{lem:E}); otherwise (b) there is no
    assignment to $b$ in $I_Y$, so $t'(b) = s'(b) = s(b) = t(b)$ (last
    step by $s \approx t$).
  \item There are no assignments to variables of type $X$ in $I_Y$,
    and
    $$
    \begin{array}{rl}
    & t'(x)\\
    =& \pstep{definition}\\
    & \alpha(s'(x))\\
    =& \pstep{no assignments to $x$ in $I^\sharp_Y$}\\
    & \alpha(s(x))\\
    =& \pstep{$\alpha$ is $X$-bijection}\\
    & t(x).
    \end{array}
    $$
  \item If $y := y'$ is in $I_Y$, then
    $$
    \begin{array}{rl}
      & t'(y)\\
      =& s'(y)\\
      =& \pstep{$y := y'$ is in $I^\sharp_Y$}\\
      & s(y')\\
      =& \pstep{$s \approx t$}\\
      & t(y').
    \end{array}
    $$
    If $\Read{y}{a}{x}$ is in $I_Y$ then
    $$
    \begin{array}{rl}
      & t'(y)\\
      =& s'(y)\\
      =& \pstep{$y := ax$ is in $I^\sharp_Y$}\\
      & s(ax)\\
      =& \pstep{$s \approx t$}\\
      & t(a[x]).\\
    \end{array}
    $$
    Otherwise, assume $y :=\ ?$ is not in $I_Y$. Therefore it's not
    in $I^\sharp_Y$, so $t'(y) = s'(y) = s(y) = t(y)$.
  \item For an array $a$ and $v \in A^*$, cases arising from the
    definition of $\Delta^*$ are:
    \begin{itemize}
    \item If $\Write{a}{x}{y}$ is in $I_Y$ and $t(x) = v$, then one of
      the following cases from the definition of $t'$ applies.
      \begin{itemize}
      \item There is an $x$ such that $t'(x) = v$. In this case $ax :=
        y$ is in the first command of $I^\sharp_Y$, and there are no
        appendages on $I^\sharp_Y$ that change $ax$. So $t'(a)(v) =
        s'(ax) = s(y) = t(y)$.
      \item Or, as $\Write{a}{x}{y}$ and $t(x) = v$, we get $t'(a)(v)
        = t(y)$ by definition.
      \end{itemize}
    \item Else, one of the following applies (taking cases from the
      definition of $t'$).
      \begin{itemize}
      \item Suppose there is some $X$-variables such that $t'(x) = v$
        (and hence $t(x) = v$ as there are no $X$-type assignments in
        $I_Y$), then notice there is no $ax := y$ in $I^\sharp_Y$.
        Also, the appendages on $I^\sharp_Y$ do not affect $ax$,
        because if they did, it would mean there exists an $x'$ such
        that $t'(x') = t'(x) = v$ and $\Write{a}{x'}{y}$ is in $I_Y$,
        and we would be in the case above. So we get $t'(a)(v) =
        s'(ax) = s(ax) = t(a[x]) = t(a)(v)$.
      \item The $\Write{a}{x}{y}$ case of the definition of $t'(a)(v)$ 
        cannot hold here, as it would be dealt with above.
      \item Otherwise $t'(a)(v) = t(a)(v)$ by definition. \mathproofbox
      \end{itemize}
    \end{itemize}
  \end{itemize}
\end{proof*}

\begin{lemma}
  \label{lem:X1}
  If $s \approx t$ and $s \Delta_{I^\sharp_X} s'$, then there exists $t'$
  such that $s' \approx t'$ and $t \Delta^*_{I_X} t'$.
\end{lemma}
\begin{proof*}
  Define a function $\alpha'$ on $s'(\TERMS_X)$ as follows:
  $$
  \begin{array}{rcll}
    \alpha'(v) &=& F(v), & \mbox{if for all $X$-type variables $x$,}\\
    &&& \mbox{ $s'(x) = v$ implies $x :=\ ?$ is in $I_X$,}\\
    &=& \alpha(v), & \mbox{otherwise,}
  \end{array}
  $$
  where $F$ is any injection from $s'(\TERMS_X)$ to $A^*
  \setminus t(\TERMS_X)$ (fresh values for $t'$ from the type $X$). We
  also restrict the range of $F$ to values which are undefined in all
  of the functions $t(a)$ for all arrays $a$. This still leaves an
  infinite number of values as the finite number of arrays are each
  finite partial functions.
  
  We need to show that $\alpha'$ is well-defined, specifically that
  $\alpha(v)$ is defined in the second case for $v$ equal to some
  $s'(x)$.  So assume there exists an $x$ such that $s'(x) = v$ and
  $x :=\ ?$ is not in $I_X$. So $x' :=\ ?$ cannot be in $I^\sharp_X$
  either.
  \begin{itemize}
  \item If there are no assignments to $x$ in $I^\sharp_X$ then
    $s'(x) = s(x)$. Therefore $v = s(x) \in s(\TERMS_X) =
    \dom(\alpha)$.
  \item If there is an assignment $x := x'$ in $I^\sharp_X$, then
    $s'(x) = s(x')$, so $v \in \dom(\alpha)$.
  \end{itemize}
  
  Now we can define $t'$ as follows:
  $$
  \begin{array}{rcl}
    t'(b) &=& s'(b),\\
    t'(x) &=& \alpha'(s'(x)),\\
    t'(y) &=& s'(y),\\
    t'(a)(v) &=& s'(ax), \mbox{ if there is such an $x$ where $t'(x)
      = v$},\\
    &=& t(a)(v), \mbox{ otherwise}.\\
  \end{array}
  $$
  Once more we need to prove that this is well-defined for the
  first case for arrays: we must have $t'(x) = t'(x')$ implies $s'(ax)
  = s'(ax')$. Notice that $\alpha'$ is injective because $\alpha$ and
  $F$ are injections with non-overlapping ranges. Therefore
  $$
  \begin{array}{rrcl}
    & t'(x) &=& t'(x')\\
    \implies& \alpha'(s'(x)) &=& \alpha'(s'(x'))\\
    \implies & && \pstep{$\alpha'$ is injective}\\
      & s'(x) &=& s'(x').
  \end{array}
  $$
  By look at the appendages on $I^\sharp_X$, it can be seen that
  $s'(x) = s'(x')$ implies $s'(ax) = s'(ax')$ when either of $x :=\ ?$
  or $x' :=\ ?$ are in $I^\sharp_X$. In more detail: if only $x :=\ ?$
  is in $I^\sharp_X$, then the first set of appendages will execute
  $ax :=\ ?$; similarly for $x'$; if both $x :=\ ?$ and $x' :=\ ?$ are
  in $I^\sharp_X$, the second set of appendages will ensure they are
  both eventually set to the least (see definition of $I^\sharp_X$ for
  this ordering) $ax_i$ such that $s'(x_i) = s'(x) = s'(x')$.
 
  When the appendages do not affect either $ax$ or $ax'$, we are left
  with the following cases:
  \begin{itemize}
  \item There are no assignments to either $x$ or $x'$ in
    $I^\sharp_X$. In which case there are no assignments to $ax$ or
    $ax'$ in $I^\sharp_X$ either, and the argument runs the same as
    the proof that $t'(x) = t'(x') \longrightarrow s'(ax) = s'(ax')$
    in the corresponding part of in Lemma \ref{lem:Y1}.
  \item There is no assignment to $x'$, but there is an assignment $x
    := x''$ in $I^\sharp_X$, in which case there is also an assignment
    $ax := ax''$ by construction of $I^\sharp_X$. We get:
    $$
    \begin{array}{rrcl}
      & t'(x) &=& t'(x')\\
      \implies & && \pstep{$x'$ not affected, $x := x''$ in $I_X$}\\
      & t(x'') &=& t(x')\\
      \implies & t(a)(t(x'')) &=& t(a)(t(x'))\\
      \implies & t(a[x'']) &=& t(a[x'])\\
      \implies & && \pstep{$s \approx t$}\\
      & s(ax'') &=& s(ax')\\
      \implies & && \pstep{$ax'$ unaffected, $ax := ax''$ in $I^\sharp_X$}\\
      & s'(ax) &=& s'(ax')
    \end{array}
    $$
  \item The cases for an assignment to only $x'$, or to both $x$ and
    $x'$, run similarly.
  \end{itemize}
  
  Notice that $\alpha'$ forms an $X$-bijection from $s$ to $t$.
  Notice further from the definition of $t'$ that $s' \approx t'$.
  
  We now wish to show that $t \Delta^*_{I_X} t'$.
  \begin{itemize}
  \item There are no boolean assignments in either $I^\sharp_X$ or
    $I_X$, so $t'(b) = s'(b) = s(b) = t(b)$.
  \item There are no assignments to variables of type $Y$ either.
  \item If $x := x'$ is in $I_X$ then
    $$
    \begin{array}{rl}
      & t'(x)\\
      & \pstep{definition of $t'$}\\
      =& \alpha'(s'(x))\\
      =& \pstep{$x := x'$ is in $I^\sharp_X$}\\
      & \alpha'(s(x'))\\
      =& \pstep{$s(x') = s'(x)$ and $x :=\ ?$ not in $I_X$}\\
      & \alpha(s(x'))\\
      =& \pstep{$s \approx t$}\\
      & t(x').
    \end{array}
    $$
    Otherwise, assume neither $x :=\ ?$ nor $x := x'$ in $I_X$.
    Therefore neither are in $I^\sharp_X$, so $t'(x) = \alpha'(s'(x))
    = \alpha'(s(x)) = \alpha(s(x)) = t(x)$, similarly to above.
  \item For an array $a$ and $v \in A^*$, taking cases from the
    definition of $\Delta^*$ for arrays.
    \begin{itemize}
    \item There is no $\Write{a}{x}{y}$ in $I_X$.
    \item Assume that there are no $X$-type variables $x$ such that $x
      :=\ ?$ is in $I_X$ and $t'(x) = v$ and $t(a)(v) = \bot$. It
      remains to show that $t'(a)(v) = t(a)(v)$.
      
      If the second case in the definition of $t'$ is invoked, then we
      get $t'(a)(v) = t(a)(v)$ immediately. So suppose instead that
      there is an $x$ where $t'(x) = v$. We will now proceed by cases
      on the command $I_X$.
      \begin{itemize}
      \item Suppose there is no assignment to $x$ in $I_X$. Then there
        are no assignments to $x$ or $ax$ in $I^\sharp_X$.  (There
        will be no assignments to $ax$ in the appendages on
        $I^\sharp_X$ because $x :=\ ?$ is not in $I$.) Starting with
        the definition of $t'$, we get $t'(a)(v) = s'(ax) = s(ax) =
        t(a[x]) = t(a)(t(x))$. Also note $t(x) = t'(x) = v$ because
        there's no assignment to $x$ in $I_X$.
      \item Suppose there is some $x'$ such that $x := x'$ is in
        $I_X$, so that $ax := ax'$ is in $I^\sharp_X$. There will be
        no assignment to $ax$ in the appendages on $I^\sharp_X$
        because $x :=\ ?$ cannot be in $I$. We get
        $$
        \begin{array}{rl}
            & t'(a)(v)\\
            =& \pstep{by definition}\\
            & s'(ax)\\
            =& \pstep{$ax := ax'$ is in $I^\sharp_X$}\\
            & s(ax')\\
            =& \pstep{$s \approx t$}\\
            & t(a[x'])\\
            =& t(a)(t(x'))\\
            =& \pstep{$x := x'$ is in $I_X$}\\
            & t(a)(t'(x))\\
            =& t(a)(v)\\
          \end{array}
          $$
        \item We are left with the case that $x :=\ ?$ is in $I_X$. We 
          will split this case further:

          (a) If there is no $x'$ such that $x' :=\ ?$ is not in $I_X$
          and $s'(x') = s'(x)$, then
          $$
          \begin{array}{rl}
            & t(a)(v)\\
            =& \pstep{how $v$ was introduced}\\
            & t(a)(t'(x))\\
            =& \pstep{definition of $t'$}\\
            & t(a)(\alpha'(s'(x)))\\
            =& \pstep{definition of $\alpha'$}\\
            & t(a)(F(s'(x)))\\
            =& \pstep{definition of $F$}\\
            & \bot.
          \end{array}
          $$
          By assumption above we are finished with this case. This
          is because the semantics of $\Delta^*$ make no requirements
          for $t'(a)(v)$ when $x :=\ ?$ is in $I_X$ and $t'(x) = v$
          and $t(a)(v) = \bot$.
          
          (b) Otherwise, there does exist an $x'$ such that $x' :=\ ?$
          is not in $I_X$ and $s'(x') = s'(x)$. Notice that $s'(x')
          \in s(\TERMS_X) = \dom(\alpha)$ because $x' :=\ ?$ is not in
          $I_X$, and we can show
          $$
          \begin{array}{rl}
            & v\\
            =& t'(x)\\
            =& \pstep{definition of $t'$}\\
            & \alpha'(s'(x))\\
            =& \pstep{definition of $\alpha'$}\\
            & \alpha(s'(x))\\
            =& \pstep{$s'(x') = s'(x)$}\\
            & \alpha(s'(x'))\\
            =& t(x').\\
          \end{array}
          $$
          As $x' :=\ ?$ is not in $I_X$, we know that $t'(a)(t(x')) =
          t(a)(t(x'))$ because of the cases we've done already.
          Therefore $t'(a)(v) = t(a)(v)$. \mathproofbox
        \end{itemize}
      \end{itemize}
  \end{itemize}
\end{proof*}

\begin{lemma}
  \label{lem:Y2}
  If $s \approx t$ and $t \Delta^*_{I_Y} t'$, then there exists $s'$ such
  that $s' \approx t'$ and $s \Delta_{I^\sharp_Y} s'$.
\end{lemma}
\begin{proof*}
  Define
  $$
  \begin{array}{rcl}
    s'(b) &=& t'(b)\\
    s'(x) &=& \alpha^{-1}(t'(x))\\
    s'(y) &=& t'(y)\\
    s'(ax) &=& t'(a[x])\\
  \end{array}
  $$
  Clearly $s' \approx t'$ (using $\alpha$ as the $X$-bijection). We now
  wish to show that $s \Delta_{I^\sharp_Y} s'$.
  \begin{itemize}
  \item For boolean variables $b$, if $b := e$ is in $I^\sharp_Y$ then
    $b := e$ appears in $I_Y$ as the only assignment to $b$. We get
    $s'(b) = t'(b) = E_t(e) = E_s(e)$ by Lemma \ref{lem:E}. Otherwise
    $s'(b) = t'(b) = t(b) = s(b)$ by $s \approx t$.
  \item There are no assignments to variables of type $X$ in
    $I^\sharp_Y$ or $I_Y$ so $s'(x) = \alpha^{-1}(t'(x)) = \alpha^{-1}(t(x)) =
    s(x)$.
  \item 
    \begin{itemize}
    \item If $y := y'$ is in $I^\sharp_Y$ then it must also be in $I_Y$,
      so $s'(y) = t'(y) = t(y') = s(y')$.
    \item if $y := ax$ is in $I^\sharp_Y$, then there must be
      $\Read{a}{x}{y}$ in $I_Y$. So $s'(y) = t'(y) = t(a[x]) = s(ax)$.
    \item else if there is no assignment to $y$ in $I^\sharp_Y$ then
      there's none in $I_Y$, so $s'(y) = t'(y) = t(y) = s(y)$.
    \end{itemize}
  \item For arrays $a$ and variables $x$ of type $X$,
    \begin{itemize}
    \item If there is an assignment $ax := y$ in the first multiple
      assignment of $I^\sharp_Y$, then the appendages on $I^\sharp_Y$
      should not affect $ax$ (see definition of $I^\sharp_Y$). Therefore
      we should have $s'(ax) = s(y)$.  It also means $\Write{a}{x}{y}$
      in $I_Y$.
        $$
      \begin{array}{rl}
        & s'(ax)\\
        =& \pstep{definition of $s'$}\\
        & t'(a[x])\\
        =& \pstep{$\Write{a}{x}{y}$ in $I_Y$}\\
        & t(y)\\
        =& \pstep{$s \approx t$}\\
        & s(y).\\
      \end{array}
        $$
    \item Now assume there is no assignment $ax := y$ in the first
      multiple assignment. Splitting cases further:
      \begin{itemize}
      \item Assume there is no $x'$ such that $ax' := y'$ is in the
        first multiple assignment in $I^\sharp_Y$, where $s(x) =
        s(x')$. This ensures that the appendages on $I^\sharp_Y$ do
        not affect $ax$, because the condition $x' = x$ is never met,
        and we should get $s'(ax) = s(ax)$. By definition of
        $I^\sharp_Y$, this means that there is no $\Write{a}{x'}{y}$
        in $I_Y$ where $t(x) = t(x')$, so $t'(a)(t(x)) = t(a)(t(x))$.
        We now get:
        $$
        \begin{array}{rl}
          & s'(ax)\\
          =& \pstep{definition of $s'$}\\
          & t'(a[x])\\
          =& t'(a)(t'(x))\\
          =& \pstep{no assignments to $x$ in $I_Y$}\\
          & t'(a)(t(x))\\
          =& \pstep{no $\Write{a}{x'}{y}$ where $t(x) = t(x')$}\\
          & \pstep{and no $x :=\ ?$ in $I_Y$}\\
          & t(a)(t(x))\\
          =& t(a[x])\\
          & \pstep{$s \approx t$}\\
          =& s(ax)\\
        \end{array}
        $$
      \item Now assume there is an $x'$ such that $ax' := y'$ is in
        the first multiple assignment in $I^\sharp_Y$, where $s(x) =
        s(x')$. This means that the appendage $x' = x \longrightarrow
        ax := ax'$ should affect $ax$, and so we need to show that
        $s'(ax) = s(y')$.
        
        From the existence of $ax' := y'$ in $I^\sharp_Y$, we deduce
        $\Write{a}{x'}{y'}$ is in $I_Y$.
        \[
        \begin{array}{rl}
          & s'(ax)\\
          =& t'(a[x])\\
          =& t'(a)(t'(x))\\
          =& \pstep{no assignments to $x$ in $I_Y$}\\
          & t'(a)(t(x))\\
          =& \pstep{$s(x) = s(x')$ and $s \approx t$}\\
          & t'(a)(t(x'))\\
          =& \pstep{$\Write{a}{x'}{y'}$ in $I_Y$}\\
          & y'. \mathproofbox
        \end{array}
        \]
      \end{itemize}
    \end{itemize}
  \end{itemize}
\end{proof*}

\begin{lemma}
  \label{lem:X2}
  If $s \approx t$ and $t \Delta^*_{I_X} t'$, then there exists $s'$ such
  that $s' \approx t'$ and $s \Delta_{I^\sharp_X} s'$.
\end{lemma}
\begin{proof*}
  Define
  \[
  \begin{array}{rcl}
    s'(b) &=& t'(b)\\
    s'(x) &=& \alpha^{-1}(t'(x))\\
    s'(y) &=& t'(y)\\
    s'(ax) &=& t'(a[x])
  \end{array}
  \]
  Clearly $s' \approx t'$. Now to show $s \Delta_{I^\sharp_X} s'$:
  \begin{itemize}
  \item No boolean assignments in either $I_X$ or $I^\sharp_X$. So
    $s'(b) = t'(b) = t(b) = s(b)$.
  \item No assignments to any variable $y$ of type $Y$ either.
  \item For each $X$-type variable $x$,
    \begin{itemize}
    \item if $x := x'$ is in $I^\sharp_X$, then it's also in $I_X$. We 
      get $s'(x) = \alpha^{-1}(t'(x)) = \alpha^{-1}(t(x')) = s(x')$;
    \item else if $x :=\ ?$ is not in $I^\sharp_X$, then it's not in
      $I_X$, so $s'(x) = s(x)$.
    \end{itemize}
  \item For each array $a$ and $X$-type variables $x$,
    \begin{itemize}
    \item Suppose there's no assignment to $ax$ in the first multiple
      assignment of $I^\sharp_X$. This means there is no assignment to
      $x$ in $I_X$, in which case $ax$ should not be affected by the
      appendages on $I^\sharp_X$ (because $x :=\ ?$ can not be in
      $I_X$). We therefore need to show $s'(ax) = s(ax)$, which can be 
      done as follows:
      \[
      \begin{array}{rl}
        & s'(ax)\\
        =& t'(a[x])\\
        =& t'(a)(t'(x))\\
        =& \pstep{no assignment to $x$ in $I_X$}\\
        & t'(a)(t(x))\\
        =& \pstep{no $\Write{a}{x}{y}$ in $I_X$}\\
        & t(a)(t(x))\\
        =& t(a[x])\\
        =& \pstep{$s \approx t$}\\
        & s(ax).\\
      \end{array}
      \]
    \item Suppose there's an assignment $ax := ax'$ in $I^\sharp_X$,
      which means there's an assignment $x := x'$ in $I_X$. Again, the
      appendages should not affect $ax$, so we expect that $s'(ax) =
      s(ax')$. The proof runs similarly to the previous case, except
      that $t'(x) = t(x')$: $s'(ax) = t'(a[x]) = t'(a)(t'(x)) =
      t'(a)(t(x')) = t(a)(t(x')) = t(a[x']) = s(ax')$.
    \item We are left with the case that $ax :=\ ?$ is in
      $I^\sharp_X$, in which case $x :=\ ?$ is in $I_X$.
      \begin{itemize}
      \item Suppose $s'(x) \neq s'(x')$ for all other variables $x'$
        of type $X$. Then non of the appendages should affect $ax$,
        and the only assignment to $ax$ is the $ax :=\ ?$. In this
        case, $\Delta$ makes no demands on the value of $s'(ax)$.
      \item Suppose $s'(x) = s'(x')$ for some variables $x'$ where $x'
        :=\ ?$ is not in $I_X$. In this case, the first set of
        appendages should ensure that the command $ax := ax'$ is
        executed.
        
        The second set of appendages should not change $ax$. For
        suppose there is another $x''$ such that $s'(x) = s'(x'')$ and
        $x'' :=\ ?$ in $I_X$, then the assignment $ax := ax''$ will
        have no effect because the first set of appendages will also
        have performed $ax'' := ax'$.
        
        We can prove $s'(ax) = s'(ax')$ as follows:
        \[
        \begin{array}{rl}
        & s'(ax)\\
        =& \pstep{definition $t'$}\\
        & t'(a[x])\\
        =& t'(a)(t'(x))\\
        =& \pstep{$s'(x) = s'(x')$ and $s' \approx t'$}\\
        & t'(a)(t'(x'))\\
        =& t'(a[x'])\\
        =& s'(ax').
        \end{array}
        \]
        We have already established that $s'(ax')$ is correct with
        respect to the definition of $\Delta$ in one of the cases
        above, so $s'(ax)$ must also be correct.
      \item Suppose $s'(x) = s'(x')$ only for variables $x'$ where $x'
        :=\ ?$ is in $I_X$. In this case, the first set of appendages
        should not change $ax$, and the second set should ensure
        $s'(ax) = s'(ax')$, although this is all we need to show
        because one of these variables is nondeterministically
        selected in the first multiple assignment in $I^\sharp_X$.
        It can be shown as follows: $s'(ax) = t'(a[x]) =
        t'(a)(t'(x)) = t'(a)(t'(x')) = t'(a[x']) = s'(ax')$. \mathproofbox
      \end{itemize}
    \end{itemize}
  \end{itemize}
\end{proof*}

\begin{proposition}
  \label{prop:bisim}
  For any program $\cal P$, and any infinite sets $A^*$ and $B^*$, the
  relation $\approx$ forms a bisimulation between $\semi{\cal
    P^\sharp}{A^*, B^*}$ and $\pfsem{\cal P}{A^*, B^*}$.
\end{proposition}
\begin{proof*}
  
  The proof is presented in three parts: first the base condition,
  followed by the two successor conditions.

\begin{enumerate}
\item Assume $s \in Q$ and $t \in Q^*$ and $s \approx t$. Note
  $$
  \begin{array}{rl}
    & s \in \ext{b} \\
    \iff & \\
    & s(b) = \true \\
    \iff & \quad \{ s \approx t \} \\
    & t(b) = \true \\
    \iff & \\
    & t \in \ext{b}^*.
  \end{array}
  $$
  So for observables $p$, we have $s \in \ext{p}$ if and
  only if $t \in \ext{p}^*$.
  
\item Take any $s, s' \in Q$ and any $t \in Q^*$ such that $s \approx t$
  and $s' \in \delta(s)$. So there exists some $e \longrightarrow
  I^\sharp$ from $\cal P$ such that $E_s(e) = \true$ and $s
  \Delta_{I^\sharp} s'$.
  
  By Lemma \ref{lem:E}, we can shown $E_t(e) =
  \true$.
  
  By construction of $I^\sharp$, we know there exists $s''$ such that
  $s \Delta_{I^\sharp_Y} s''$ and $s'' \Delta_{I^\sharp_X} s$.
  
  By Lemma \ref{lem:Y1}, we know there exists $t''$ such that $t
  \Delta^*_{I_Y} t''$ and $s'' \approx t''$. By Lemma \ref{lem:X1}, we know
  there exists $t'$ such that $t'' \Delta^*_{I_X} t'$ and $s' \approx t'$.
  By Lemma \ref{lem:comp}, $t \Delta^*_I t'$.
\item This case runs symmetrically to the above case. Use Lemma
  \ref{lem:comp} to show $t \Delta^*_I t'$ is equivalent to $t
  \Delta^*_{I_Y} t''$ and $t'' \Delta^*_{I_X} t'$ for some $t'' \in
  Q^*$. Use Lemmas \ref{lem:Y2} and \ref{lem:X2} instead where
  appropriate, and the last step should be replaced with the
  observation that $s \Delta_{I^\sharp_Y} s''$ and $s''
  \Delta_{I^\sharp_X} s'$ implies
  \[
  s \Delta_{I^\sharp_Y : \true \longrightarrow I^\sharp_X} s'
  \]
  by definition of $:$ the append operator. \mathproofbox
\end{enumerate}
\end{proof*}

\subsection{Main theorem}

\label{sub:conclusion}

We are now ready to present our first main result: that the
$\mu$-calculus model-checking problem is decidable for the class of
systems generated from programs using partial-functions semantics and
infinite instantiations for $X$ and $Y$.

\begin{theorem}
  \label{thm:infinite}
  Given
  \begin{itemize}
  \item a program $\cal P$,
  \item a boolean variable $b_0$ of $\cal P$,
  \item a $\mu$-calculus formula $\varphi$ over the boolean variables
    of $\cal P$,
  \end{itemize}
  for any infinite sets $A^*$ and $B^*$ (over which equality
  is decidable), the model-checking problem $\pfsem{P}{A^*,
    B^*}, b_0 \models \varphi$ is decidable. Moreover, the answer
  is independent of which infinite sets $A^*$ and $B^*$ are
  used.
\end{theorem}
\begin{proof}
  The array-free abstraction ${\cal P}^\sharp$ of $\cal P$ is a
  data-independent program without arrays, and the array-consistency
  formula $\Sigma$ from Definition \ref{def:approx} uses only equality
  on the variables of ${\cal P}^\sharp$.  Therefore, it is possible to
  generate a finite transition system $M$ which has the same
  observables as, and is bisimulation-equivalent to, the transition
  system $\semi{\cal P^\sharp}{A^*, B^*}$ using the algorithm in
  \cite{NK00} with $\Sigma$ as the initial condition\footnote{ The
    syntax of programs used in \cite{NK00} is almost identical to
    ours. The semantics are given in terms of weakest liberal
    precondition laws, which can be related to our operational
    semantics in the standard way \cite{Hoa69}.  The append operator
    used here is easily integrated into \cite{NK00} using the weakest
    liberal precondition law $$\{\mathit{wlp}_{I_1} (\psi \wedge \neg
    e) \vee \mathit{wlp}_{I_1} (\mathit{wlp}_{I_2} (\psi) \wedge e) \}
    \quad I_1 : e \longrightarrow I_2 \quad \{ \psi \}.$$}.
  
  Also note that states related by some bisimulation have exactly the
  same true $\mu$-calculus formulas \cite{BCG88}.
  
  Using these facts we proceed as follows:
  \[
  \begin{array}{rl}
    & \pfsem{\cal P}{A^*, B^*}, b_0 \models \varphi\\
    \iff
    & \forall t \in \ext{b_0}^* \cdot \pfsem{\cal P}{A^*, B^*}, t
    \models \varphi\\
    \iff & \pstep{Proposition \ref{prop:bisim}
      and Definition \ref{def:approx}}\\
    & \forall s \in \ext{b_0}^\sharp \cdot \Sigma(s) \implies
    \semi{\cal P^\sharp}{A^*, B^*}, s \models \varphi\\
    \iff & \pstep{\cite{NK00}}\\
    & \forall u \in \ext{b_0} \cdot M, u \models \varphi\\
    \iff
    & M, b_0 \models \varphi.
  \end{array}
  \]
  Hence the problem can be solved by $\mu$-calculus finite-model
  checking, for example \cite{BC+92}.
  
  The independence of $A^*$ and $B^*$ comes from the fact that these
  sets are not actually used by \cite{NK00} in the construction of the
  finite transition system $M$.
\end{proof}

The above proof suggests the following procedure for model checking
data-independent systems with arrays.  Suppose a program $\cal P$ has
$n_b$ boolean variables, $n_x$ variables of type $X$, $n_y$ variables
of type $Y$, $n_a$ array variables, and $n_i$ guarded commands.
\begin{enumerate}
\item {\bf Translate $\cal P$ to its array-free abstraction $\cal
    P^\sharp$ using the procedure in Section \ref{sub:translation}.}
  The translation procedure will produce a program with the same
  number of boolean variables, $n_x$ variables of type $X$, $n_y + n_a
  n_x$ variables of type $Y$, and no array variables. The complexity
  of commands is increased due to the append operator and we will
  count each one as a separate command. There are a maximum of
  $\frac{1}{4} n_a n_x^2$ appendages added onto each $I^\sharp_Y$, and
  a maximum of $\frac{1}{2} n_a n_x^2$ added onto each $I^\sharp_X$.
  The total number of guarded commands in $\cal P^\sharp$ could be as
  high as
  \[
  n_i ( \frac{3 n_a n_x^2}{4} + 2 ).
  \]
  As this translation can be done instruction by instruction, its time
  complexity is equivalent to the above bound on the number of guarded
  commands that may appear in $\cal P^\sharp$.
\item {\bf Translate $\cal P^\sharp$, under the initial condition of
    the array-consistency formula $\Sigma$, to the finite state
    transition system $M$ using the syntactic transformation procedure
    in \cite{NK00}.}  This procedure would generate at most $n_x^2 +
  (n_y + n_a n_x)^2 + n_b$ predicates, and therefore would terminate
  in at most that number of steps\footnote{The complexity of each step
    of the algorithm in \cite{NK00} is not given, although it appears
    that the total complexity of the algorithm is at least in $\Omega(
    p^2 l )$, where $p$ is the number of predicates generated and $l$
    is the number of guarded commands.}. The number of states in $M$
  would be at most
  \[
  n_b  n_x^{n_x} (n_y + n_a n_x)^{(n_y + n_a n_x)}.
  \]
\item {\bf Model check $M$ using any finite-model-checking algorithm,
    eg \cite{BC+92}.} Finite-model checking of the $\mu$-calculus in
  general is {\em EXP\-SPACE} in the size of the model.
\end{enumerate}

Instead of steps 2 and 3 above, there are other ways we might solve
$$\semi{P^\sharp}{A^*, B^*}, b_0 \models \varphi.$$
One way would be
to use a finite instantiation theorem \cite{LN00}.  A more efficient
way would be to design a region algebra and use the model-checking
algorithm in \cite{HM00}.  However, the syntactic translation in
\cite{NK00} first generates a bisimulation-equivalent program with
just boolean variables, and orthogonal techniques could be applied to
that program before using it to generate the transition system $M$.

\begin{figure}
{\footnotesize
\begin{verbatim}
INITIALLY:
    addrBus = testData => (mem1_addrBus = mem1_testData /\
            mem2_addrBus = mem2_testData /\ mem3_addrBus = mem3_testData)
\end{verbatim}}
\caption{Initial condition for array-free abstraction of the
  fault-tolerant memory composed with specification.
\label{fig:singleCacheInitially}}
\end{figure}

\begin{example}
  \label{ex:singleCacheInfinite}
  We will now begin to show how to check that the program in Example
  \ref{ex:singleCache} satisfies its specification.

 Following the steps outlined above:
\begin{enumerate}
\item The translation of the program $\cal P$ to its array-free
  abstraction $\cal P^\sharp$ is shown in Figure
  \ref{fig:singleCacheTransl}.
\item The array-free abstraction $\cal P^\sharp$, together with the
  initial condition shown in Figure \ref{fig:singleCacheInitially}, can be
  converted to a finite state transition system $M$ as described in
  \cite{NK00}.
\item We can now perform the check $M, b_0 \models \nu h: \varphi$, where
  $\varphi$ is $\forallnext (\overline{b_E} \wedge h)$.
\end{enumerate}
The proof of Theorem \ref{thm:infinite} tells us that the answer given
by this check will be equivalent to the answer of $\pfsem{\cal
  P}{A^*, B^*}, b_0 \models \varphi$ for any infinite sets
$A^*$ and $B^*$.
\end{example}

\section{Finite arrays}

\label{sect:finite}

In this section we present results about the class of programs with
arbitrary non-empty finite sets as instantiations for their types. By
showing the relationship between one transition system generated using
infinite sets and all systems generated using finite sets, we are able
to deduce how fragments of the $\mu$-calculus are preserved between
them.

\begin{proposition}
\label{prop:simulation}
  For any non-empty finite sets $A$ and $B$, and infinite respective
  supersets $A^*$ and $B^*$, there exists a total simulation of
  $\semi{\cal P}{A, B}$ by $\pfsem{\cal P}{A^*, B^*}$.
\end{proposition}
\begin{proof*}
Let
$$
\begin{array}{rlcl}
& \semi{\cal P}{A, B} &=& (Q, \delta, \ext{\cdot}, P) \\
\mbox{ and } & \pfsem{\cal P}{A^*, B^*} &=&
  (Q^*, \delta^*, \ext{\cdot}^*,
P).
\end{array}
$$

Define a total relation $\lhd \subseteq Q \times Q^*$ as $s \lhd t$ if and
only if $s$ and $t$ are identical, except that for arrays $a$,
we have $t(a)(v)$ is equal to $s(a)(v)$ if $v \in A$, and $\bot$ if $v
\in A^* \setminus A$.

For the first condition of simulation, observe that
$$
\begin{array}{rl}
& s \in \ext{b} \\
\iff & \\
& s(b) = \true \\
\iff & \quad \{ s \lhd t \} \\
& t(b) = \true \\
\iff & \\
& t \in \ext{b}^*.
\end{array}
$$
So for observables $p$, we have $s \in \ext{p}$ if
and only if $t \in \ext{p}^*$.

For the second condition, assume that $s \lhd t$ and $s' \in
\delta(s)$. We need to show that there exists $t' \in Q^*$ such that
$t' \in \delta(t)$ and $s' \lhd t'$.

Define $t'$ by $s' \lhd t'$. As $s' \in \delta(s)$, there must exist a
guarded command $e \longrightarrow I$ in ${\cal P}$ such that
$E_{s}(e) = \true$ and $s \Delta_I s'$.
\begin{itemize}
\item $E_{t}(e) = E_{s}(e)$ by (an easy variation of) Lemma \ref{lem:E}.
\item It remains to show $t \Delta^*_I t'$. We do only the case for
  arrays.
  \begin{itemize}
  \item for each array variable $a$, and for each $v \in A^*$,\\
    \begin{tabular}{l}
      if there are $x$ and $y$ variables such that \\
      \quad  $\Write{a}{x}{y} \in I$ and $t(x) = v$, then \\
      \quad  $t'(a)(v) = s'(a)(v) = s(y) = t(y)$. \\
      else either $t'(a)(v) = \bot = t(a)(v)$, \\
      \quad or $t'(a)(v) = s'(a)(v) = s(a)(v) = t(a)(v)$. \mathproofbox
    \end{tabular}
  \end{itemize}
\end{itemize}
\end{proof*}

\begin{proposition}
\label{prop:traces}
For any infinite sets $A^*$ and $B^*$, if $\pi$ is a trace
of $\pfsem{\cal P}{A^*, B^*}$, then there exist non-empty
finite respective subsets $A$ and $B$ such that $\pi$ is a trace of
$\semi{\cal P}{A, B}$.
\end{proposition}
\begin{proof}
  Let
  $$
  \pfsem{\cal P}{A^*, B^*} = (Q^*, \delta^*,
  \ext{\cdot}^*, P).
  $$
  If $\pi$ is a trace of $\pfsem{\cal P}{A^*, B^*}$, then
  there exists a sequence $t_1 t_2 ... t_l$ of states from $Q^*$ such
  that $t_{i+1} \in \delta^*(t_i)$ for $i = 1...l-1$, and $t_i \in
  \ext{ \pi(i) }$ for $i = 1...l$.
  
  As the functions representing arrays in these states are finite
  partial functions, they contain only finite subsets $A$ and $B$ of
  $A^*$ and $B^*$. We can now form the transition system
  $$
  \semi{\cal P}{A, B} = (Q, \delta, \ext{\cdot}, P).
  $$
  
  Form a state $s_l \in Q$ from $t_l$ as follows. Extending the
  partial functions in $t_l$ to total functions on $A$ by picking any
  $B$ values for the undefined locations. Now, working backwards from
  $i = l-1$ down to $i = 1$, form states $s_i \in Q$ by extending the
  partial functions in $t_i$ to total functions using the same values
  used for $s_{i+1}$.

  Formally,
  $$
  \begin{array}{rcl}
  s_i(b) &=& t_i(b), \\
  s_i(z) &=& t_i(z), \\
  s_i(a)(v) &=& t_i(a)(v), \mbox{ if defined, else} \\
  &=& \mbox{ anything, if } i = l, \\
  &=& s_{i+1}(a)(v), \mbox{ otherwise,} \\
  \end{array}
  $$
  for boolean variables $b$, data variables $z$, arrays variables
  $a$ and values $v$ from $A$.

  We now wish to show that $s_{i+1} \in \delta(s_i)$ for $i =
  1...l-1$. As $t_{i+1} \in \delta^*(t_i)$, there must exist a
  guarded command $e \longrightarrow I$ in ${\cal P}$ such that 
  $E_{t_i}(e) = \true$ and $t_i \Delta_I t_{i+1}$.
  \begin{itemize}
  \item $E_{t}(e) = E_{s}(e)$ by (an easy variation of) Lemma \ref{lem:E}.
  \item It remains to show $s_i \Delta^*_I s_{i+1}$. We do only the
    case for arrays.
    \begin{itemize}
    \item For each array $a$ and each $v \in A$,
      \begin{itemize}
      \item If $\Write{a}{x}{y} \in I$ and $s(x) = v$, then
        $s_{i+1}(a)(v) = t_{i+1}(a)(v)$, which must be defined because
        $t_i(x) = v$. From $t_i \Delta^*_I t_{i+1}$ we know
        $t_{i+1}(a)(v) = t_i(y)$, and by definition $s_i(y) = t_i(y)$.
        So $s_{i+1}(a)(v) = s_i(y)$.
      \item Else, if $t_{i+1}(a)(v)$ is defined anyway, $s_{i+1}(a)(v)
        = t_{i+1}(a)(v)$. Two cases arise from the definition of
        $\Delta^*$.
        \begin{itemize}
        \item Either there is an $x :=\ ?$ in $I$ and $t_{i+1}(x) = v$
          and $t_i(a)(v) = \bot$. The last of these means that
          $s_i(a)(v) = s_{i+1}(a)(v)$ by definition.
        \item Or $t_{i+1}(a)(v) = t_i(a)(v)$. Whether this is a value
          from $A$ or it is $\bot$, by definition $s_i(a)(v) =
          s_{i+1}(a)(v)$.
        \end{itemize}
      \end{itemize}
    \end{itemize}
    This is enough to show $s_i \Delta^*_I s_{i+1}$ for the arrays case.
  \end{itemize}
  This shows that the sequence $s_1...s_l$ is an execution sequence in
  $\semi{\cal P}{A, B}$. Notice also that $\ext{s_i} = \ext{t_i}$
  because they are equivalent at the boolean variables, so $\pi$ is a
  trace of $\semi{\cal P}{A, B}$.
\end{proof}

\begin{definition}
The open formulas of the logic $L^\infty_4$ over a set of observables
$P$ are generated by the grammar $\psi$:
\[
\psi ::= \bigvee_i \psi'_i\ ; \quad \psi' ::= p \mid h \mid \existsnext \psi',
\]
for $p \in P$ and variables $h$, where $\bigvee_i \psi'_i$ represents
any countable disjunction of formulas from the grammar $\psi'$.
\end{definition}

Given a transition system ${\cal S} = (Q, \delta, \ext{\cdot}, P)$
and a mapping from the variables to sets of states $\Epsilon$, any
open formula $\varphi$ of $L^\infty_4$ over $P$ defines a set
$\syns{\varphi}{\Epsilon} \subseteq Q$ of states:
$$
\begin{array}{l}
\syns{p}{\Epsilon} = \ext{p} \\
\syns{h}{\Epsilon} = \Epsilon(h) \\
\syns{\existsnext \psi}{\Epsilon} = \{ s \in Q
\mid \exists s' \in \delta(s) : s' \in
\syns{\psi}{\Epsilon} \} \\
\syns{\bigvee_i \psi_i}{\Epsilon} =
\bigcup_i \syns{\psi_i}{\Epsilon}.
\end{array}
$$

\begin{proposition}
\label{prop:l4}
Any closed $\mu$-calculus formula $\varphi \in L^\mu_4$ is semantically
equivalent to a closed formula $\psi \in L^\infty_4$.
\end{proposition}
\begin{proof*}
Define a function $F$ from open $L^\mu_4$ formulas to open $L^\infty_4$
formulas. For ease of presentation, we will write disjunction as sets
in the target language.
\[
\begin{array}{rcl}
F(p) &=& \{ p \} \\
F(h) &=& \{ h \} \\
F(\varphi_1 \vee \varphi_2) &=& F(\varphi_1) \cup F(\varphi_2) \\
F(\existsnext \varphi) &=& \mathrm{map} \ \existsnext \ F(\varphi) \\
F(\mu h : \varphi) &=& \bigcup_{i \in \mathbbm{N}} \psi_i \\
&& \begin{array}{rcl}
   \mbox{where } \psi_0 &=& \{ \} \\
     \psi_{i+1} &=& N( \subst{F(\varphi)}{\psi_i}{h} ).
   \end{array}
\end{array}
\]
The function $N$ is a function which normalises formulas from the
grammar
\[
\psi'' ::= p \mid h \mid \bigvee_i \psi''_i \mid \existsnext \psi''
\]
to formulas from $L^\infty_4$, and is defined as follows:
\[
\begin{array}{rcl}
  N(p) &=& \{ p \} \\
  N(h) &=& \{ h \} \\
  N( \bigvee_i \psi_i ) &=& \bigcup_i \psi_i \\
  N( \existsnext \psi) &=& \{ \}, \mbox{ if } N(\psi) = \{ \} \\
    && \mathrm{map} \ \existsnext \ N(\psi), \mbox{ otherwise.}
\end{array}
\]
Note that these functions are well defined as their definitions are
inductive.

It can be shown by structural induction that the function $N$
preserves the semantics of formulas because $\existsnext$ distributes
over disjunction and $\existsnext \false$ is equivalent to $\false$.

It can further be shown that $F$ also preserves the semantics of
formulas. We will do only the $\mu$ case, using a result from
\cite{Sti92} due to the fixed-point theorem for continuous functions
over complete partial orders which allows us to replace occurrences of
$\mu$ in formulas with infinite disjunction.
\[
\begin{array}{rl}
& \syns{\mu h : \varphi}{\Epsilon} \\
=& \pstep{ \cite{Sti92} } \\
& \begin{array}{l}
  \syns{\bigcup_{i \in \mathbbm{N}} \psi_i}{\Epsilon} \\
  \begin{array}{rcl}
                 \mbox{ where } \psi_0 &=& \{\} \\
                 \psi_{i+1} &=&  \subst{\varphi}{\psi_i}{h} \\
                 &=& \pstep{ induction hypothesis} \\
                 && \subst{F(\varphi)}{\psi_i}{h} \\
                 &=& \pstep{ $N$ preserves semantics} \\
                 && N( \subst{F(\varphi)}{\psi_i}{h} ) \\
  \end{array}
\end{array} \\
=& \pstep { definition of $F$ } \\
& \syns{ F( \mu h : \varphi) }{\Epsilon} \mathproofbox
\end{array}
\]
\end{proof*}

We now present our second main result, which relates the
model-checking procedure for systems with infinite arrays presented in
Section \ref{sect:infinite} to the parameterised model-checking
problem for systems with finite arrays.
\begin{theorem}
  \label{thm:finite}
  For
  \begin{itemize}
  \item a program $\cal P$,
  \item a boolean variable $b_0$ of $\cal P$,
  \item a $\mu$-calculus formula $\overline{\varphi}$ over the boolean variables
    of $\cal P$,
  \item infinite sets $A^*$ and $B^*$ (over which equality
    is decidable),
  \end{itemize}
  we have, for $A$ and $B$ necessarily finite non-empty subsets of
  $A^*$ and $B^*$ respectively:
  \begin{enumerate}
  \item For $\overline{\varphi}$ in the universal fragment of the $\mu$-calculus
    $\overline{L^\mu_2}$,
    \[
    \pfsem{\cal P}{A^*, B^*}, b_0 \models \overline{\varphi} \quad
    \Longrightarrow \quad \forall A, B \cdot \semi{\cal P}{A, B}, b_0
    \models \overline{\varphi}.
    \]
  \item For $\overline{\varphi}$ in the universal disjunction-free fragment of
    the $\mu$-calculus $\overline{L^\mu_4}$,
    \[
    \pfsem{\cal P}{A^*, B^*}, b_0 \models \overline{\varphi}
    \quad \Longleftrightarrow \quad
    \forall A, B \cdot \semi{\cal P}{A, B}, b_0 \models \overline{\varphi}.
    \]
  \end{enumerate}
\end{theorem}
\begin{proof}
  For Part 1, Notice:
  \[
  \begin{array}{rrcl}
    & \pfsem{\cal P}{A^*, B^*}, b_0 \models \overline{\varphi}
    & \Longrightarrow &
    \forall A, B \cdot \semi{\cal P}{A, B}, b_0 \models \overline{\varphi}\\
    \iff &&& \pstep{definition of $\models$}\\
    & \forall t \in \ext{b_0}^* \cdot \pfsem{\cal P}{A^*,
      B^*}, t \models \overline{\varphi}
    & \Longrightarrow &
    \forall A, B \cdot \forall s \in \ext{b_0} \cdot \semi{\cal P}{A,
      B}, s \models \overline{\varphi}.\\
  \end{array}
  \]
  So assuming the left-hand side, take any finite non-empty subsets
  $A$ and $B$ of $A^*$ and $B^*$ respectively, and any state
  $s \in \ext{b_0}$.
  
  By Proposition \ref{prop:simulation}, there exists a total simulation of
  $\semi{\cal P}{A, B}$ by $\pfsem{\cal P}{A^*, B^*}$, so there must
  exist a state $t \in \ext{b_0}^*$ such that $t$ simulates $s$. By
  \cite{GL94}\footnote{For any $\overline{L^\mu_2}$ formula
    $\overline{\varphi}$, if $t$ simulates $s$ then $M, t \models
    \overline{\varphi}$ implies $M, s \models \overline{\varphi}$.},
  we can conclude the right-hand side.
  
  The forward direction of Part 2 follows from the first result
  because $\overline{L^\mu_4} \subseteq \overline{L^\mu_2}$.  For the
  reverse direction, notice, for $\varphi \in L^\mu_4$ the dual formula
  of $\overline{\varphi} \in \overline{L^\mu_4}$:
  \[
  \begin{array}{rrcl}
    &
    \pfsem{\cal P}{A^*, B^*}, b_0 \models \overline{\varphi} &
    \Longleftarrow &
    \forall A, B \cdot \semi{\cal P}{A, B}, b_0 \models \overline{\varphi}\\
    \iff &&& \pstep{definition of $\models$}\\
    &
    \forall t \in \ext{b_0}^* \cdot \pfsem{\cal P}{A^*,
      B^*}, t \models \overline{\varphi} &
    \Longleftarrow &
    \forall A, B \cdot \forall s \in \ext{b_0} \cdot \semi{\cal P}{A,
      B}, s \models \overline{\varphi}\\
    \iff &&& \pstep{definition of $\overline{L^\mu_i}$}\\
    &
    \forall t \in \ext{b_0}^* \cdot \pfsem{\cal P}{A^*,
      B^*}, t \not \models \varphi &
    \Longleftarrow &
    \forall A, B \cdot \forall s \in \ext{b_0} \cdot \semi{\cal P}{A,
      B}, s \not \models \varphi\\
    \iff &&& \pstep{contrapositive}\\
    &
    \exists t \in \ext{b_0}^* \cdot \pfsem{\cal P}{A^*,
      B^*}, t \models \varphi &
    \Longrightarrow &
    \exists A, B \cdot \exists s \in \ext{b_0} \cdot \semi{\cal P}{A,
      B}, s \models \varphi.
  \end{array}
  \]
  We will prove this equivalent statement instead.
  
  Suppose there exists a state $t \in \ext{b_0}^*$ such that
  $\pfsem{\cal P}{A^*, B^*}, t \models \varphi$. Using Proposition
  \ref{prop:l4}, it can be seen that $\varphi$ is semantically
  equivalent to a formula $\psi$, which is the infinite
  disjunction of formulas in the form $( \existsnext )^i \, b$.

  As $\pfsem{\cal P}{A^*, B^*}, t \models \varphi$ by
  assumption, it must satisfy at least one of the disjuncts of $\psi$
  in the form $( \existsnext )^i b$. That means there is a trace
  $\pi$ of $\pfsem{\cal P}{A^*, B^*}$ such that $\pi(1) =
  \ext{b_0}$ and $\pi(i) = \ext{b}$.
  
  By Proposition \ref{prop:traces}, $\pi$ is also a trace of
  $\semi{\cal P}{A, B}$ for some finite non-empty subsets $A$ and $B$
  of $A^*$ and $B^*$ respectively.  Therefore, there exists
  some $s \in \ext{b_0}$ such that $\semi{\cal P}{A, B}, s \models (
  \existsnext )^i b$, and hence $\semi{\cal P}{A, B}, s \models
  \varphi$.
\end{proof}

\begin{example}
  \label{ex:singleCacheFinite}
  We now show how to check that the program in Example
  \ref{ex:singleCache} satisfies its specification for all finite
  non-empty sets $A$ and $B$ as instances of \code{ADDR} and
  \code{DATA}, carrying on directly from Example
  \ref{ex:singleCacheInfinite}.
  
  We have shown already how to solve $\pfsem{\cal P}{A^*,
    B^*}, b_0 \models \varphi$, where $\varphi$ is $\forallnext
  (\overline{b_E} \wedge h)$, for any infinite sets $A^*$ and
  $B^*$. Because $\varphi$ is an $\overline{L^\mu_4}$ formula,
  Theorem \ref{thm:finite} further shows us that this answer is
  equivalent to the answer of $\semi{\cal P}{A, B}, b_0 \models
  \varphi$ for all non-empty finite sets $A$ and $B$. This is the
  original specification that we decided the program should satisfy
  back in Example \ref{ex:singleCache}.
\end{example}

\begin{example}
  We have checked the running example in this paper using the model
  checker Mur$\phi$ \cite{DDH92}, which accepts UNITY-like programs as
  input and performs reachability analysis on them.
  
  We used finite instantiation theorems \cite{LN00} to show that it
  was necessary to check all sizes of \verb$ADDR$ and \verb$DATA$ less
  than and equal to 2 and 11 respectively, in order to show that the
  program works for any type instantiation. We also declared these
  types as ``scalarsets'' \cite{ID96}, so that Mur$\phi$ only checks a
  representative state from each set of symmetry equivalent states.
  The property $\varphi$ is actually a non-reachability property, and
  so Mur$\phi$ could be used to check it.

  The tool reported that the state was not reachable. Using the
  theorems as explained in Examples \ref{ex:singleCacheInfinite} and
  \ref{ex:singleCacheFinite}, this shows that the program in Figure
  \ref{fig:singleCache1} does in fact satisfy its specification that a
  read from an arbitrary location will always return the value of the
  last write to that location, provided there has been one, for all
  sizes of memory and for all types of data values.

  
  
\end{example}

\section{Conclusions}

\label{sect:conclusion}

In this paper, we have considered the class of programs
data-independent with equality with respect to two distinct type
variables $X$ and $Y$, which may also use arrays indexed by values of
type $X$ and storing values from the type $Y$.

We have shown that there is a procedure for the parameterised
model-checking problem of the universal fragment of the
$\mu$-calculus, such that it always terminates, but may give false
negatives.  We have also shown that the parameterised model-checking
problem of the universal disjunction-free fragment of the
$\mu$-calculus is decidable.

These results were obtained using, as an abstraction, programs with
any infinite instances of $X$ and $Y$ where arrays are modelled by
partial functions: it was shown that the $\mu$-calculus model-checking
problem is decidable for the resulting transition systems. A method
for doing this was presented, which uses a translation to
bisimulation-equivalent data-independent programs without arrays for
which the $\mu$-calculus model-checking problem is already known to be
decidable.

This procedure was demonstrated on a fault-tolerant interface over a
set of unreliable memories. It was shown how one could check whether
the system satisfies the property that a read at an address always
returns the value of the last write to that address until a particular
number of faults occur, independently of the size of the memory and of
the type of storable data values.

We have extended the result in \cite{HIB97} by allowing many arrays
instead of just one, and also by strengthening the model checking
decidability result from linear-time temporal logic to the
$\mu$-calculus. We have clarified a technique used in \cite{McM99} by
developing decidability results for a subclass of the programs
considered there.

Related work \cite{RL01} includes the addition of a reset operation
which sets every element of an array to a particular value. There, it
is shown that adding reset to the language used in this paper makes
even reachability undecidable for programs with at least two arrays.
However, useful decidability results for reachability are obtained in
the case where the content type of the array is finite and fixed.

Work in progress and future work include investigating the effect on
these results of adding more array operations to the programs, for
example array assignment, as well as generalising the language to have
many types and multi-dimensional arrays.  Another direction for
further work is investigating the applicability of this work to model
checking memory systems such as single processor caches \cite{PH94}
and cache-coherence protocols \cite{Qad01}, as well as parameterised
networks \cite{CR00}.


\bibliography{tom}

\begin{thebibliography}{}

\bibitem[\protect\citeauthoryear{Adve and Gharachorloo}{Adve and
  Gharachorloo}{1996}]{AG96}
{\sc Adve, S.} {\sc and} {\sc Gharachorloo, K.} 1996.
\newblock Shared memory consistency models: a tutorial.
\newblock {\em Computer\/}~{\em 29,\/}~12 (Dec.), 66--76.

\bibitem[\protect\citeauthoryear{Alur and Henzinger}{Alur and
  Henzinger}{1998}]{AH98}
{\sc Alur, R.} {\sc and} {\sc Henzinger, T.} 1998.
\newblock Computer-aided verification: An introduction to model building and
  model checking for concurrent systems.
\newblock Draft.

\bibitem[\protect\citeauthoryear{Browne, Clarke, and Gr{\"u}mberg}{Browne
  et~al\mbox{.}}{1988}]{BCG88}
{\sc Browne, M.}, {\sc Clarke, E.}, {\sc and} {\sc Gr{\"u}mberg, O.} 1988.
\newblock Characterizing finite {K}ripke structures in propositional temporal
  logic.
\newblock {\em Theoretical Computer Science\/}~{\em 59}, 115--131.

\bibitem[\protect\citeauthoryear{Burch, Clarke, McMillan, Dill, and
  Hwang}{Burch et~al\mbox{.}}{1992}]{BC+92}
{\sc Burch, J.}, {\sc Clarke, E.}, {\sc McMillan, K.}, {\sc Dill, D.}, {\sc
  and} {\sc Hwang, L.} 1992.
\newblock Symbolic model checking: $10^{20}$ states and beyond.
\newblock {\em Information and Computation\/}~{\em 98,\/}~2 (June), 142--170.

\bibitem[\protect\citeauthoryear{Chandy and Misra}{Chandy and
  Misra}{1988}]{CM88}
{\sc Chandy, K.} {\sc and} {\sc Misra, J.} 1988.
\newblock {\em Parallel Program Design: A Foundation}.
\newblock Addison Wesley Publishing Company, Inc., Reading, Massachusetts.

\bibitem[\protect\citeauthoryear{Creese and Roscoe.}{Creese and
  Roscoe.}{2000}]{CR00}
{\sc Creese, S.} {\sc and} {\sc Roscoe., A.} 2000.
\newblock Data independent induction over structured networks.
\newblock In {\em International Conference on Parallel and Distributed
  Processing Techniques and Applications}. CSREA Press, Las Vegas, Nevada, USA.
\newblock \url*http://web.comlab.ox.ac.uk/oucl/research/areas/concurrency*.

\bibitem[\protect\citeauthoryear{Dill, Drexler, Hu, and Yang}{Dill
  et~al\mbox{.}}{1992}]{DDH92}
{\sc Dill, D.}, {\sc Drexler, A.}, {\sc Hu, A.}, {\sc and} {\sc Yang, C.} 1992.
\newblock Protocol verification as a hardware design aid.
\newblock In {\em Proceedings of the {IEEE} International Conference on
  Computer Design}. {IEEE} Computer Society, Cambridge, MA, USA, 522--525.

\bibitem[\protect\citeauthoryear{Finkel and Schnoebelen}{Finkel and
  Schnoebelen}{2001}]{FS01}
{\sc Finkel, A.} {\sc and} {\sc Schnoebelen, P.} 2001.
\newblock Well-structured transition systems everywhere!
\newblock {\em Theoretical Computer Science\/}~{\em 256,\/}~1--2, 63--92.

\bibitem[\protect\citeauthoryear{Grumberg and Long}{Grumberg and
  Long}{1994}]{GL94}
{\sc Grumberg, O.} {\sc and} {\sc Long, D.} 1994.
\newblock Model checking and modular verification.
\newblock {\em {ACM} Transactions on Programming Languages and Systems\/}~{\em
  16,\/}~3 (May), 843--871.

\bibitem[\protect\citeauthoryear{Henzinger and Majumdar}{Henzinger and
  Majumdar}{2000}]{HM00}
{\sc Henzinger, T.} {\sc and} {\sc Majumdar, R.} 2000.
\newblock A classification of symbolic transition systems.
\newblock In {\em Proceedings of the 17th International Symposium on
  Theoretical Aspects of Computer Science}. Lecture Notes in Computer Science.
  Springer-Verlag, Lille, France, 13--34.

\bibitem[\protect\citeauthoryear{Henzinger, Qadeer, and Rajamani}{Henzinger
  et~al\mbox{.}}{1999}]{HQR99}
{\sc Henzinger, T.}, {\sc Qadeer, S.}, {\sc and} {\sc Rajamani, S.} 1999.
\newblock Verifying sequential consistency on shared-memory multiprocessor
  systems.
\newblock In {\em Proceedings of the 11th International Conference on Computer
  Aided Verification}. Lecture Notes in Computer Science, vol. 1633.
  Springer-Verlag, Trento, Italy, 301--315.

\bibitem[\protect\citeauthoryear{Hoare}{Hoare}{1969}]{Hoa69}
{\sc Hoare, C.} 1969.
\newblock An axiomatic basis for computer programming.
\newblock {\em Communications of the {ACM}\/}~{\em 12,\/}~10, 576--580.

\bibitem[\protect\citeauthoryear{Hojati, Dill, and Brayton}{Hojati
  et~al\mbox{.}}{1997}]{HDB97}
{\sc Hojati, R.}, {\sc Dill, D.}, {\sc and} {\sc Brayton, R.} 1997.
\newblock Verifying linear temporal properties of data insensitive controllers
  using finite instantiations.
\newblock In {\em Proceedings of the 13th {IFIP} International Conference on
  Computer Hardware Description Languages and their Applications}. Toledo,
  Spain.

\bibitem[\protect\citeauthoryear{Hojati, Isles, and Brayton}{Hojati
  et~al\mbox{.}}{1997}]{HIB97}
{\sc Hojati, R.}, {\sc Isles, A.}, {\sc and} {\sc Brayton, R.} 1997.
\newblock Automatic state reduction techniques for hardware systems modelled
  using uninterpreted functions and infinite memory.
\newblock In {\em Proceedings of the {IEEE} International High Level Design
  Validation and Test Workshop}. Oakland, California.

\bibitem[\protect\citeauthoryear{Ip and Dill}{Ip and Dill}{1996}]{ID96}
{\sc Ip, C.} {\sc and} {\sc Dill, D.} 1996.
\newblock Better verification through symmetry.
\newblock In {\em Symmetry in Automatic Verification}, {E.~Emerson}, Ed. Formal
  Methods in System Design, vol. 9 (1--2). Kluwer, 41--75.

\bibitem[\protect\citeauthoryear{Lazi\'c and Nowak}{Lazi\'c and
  Nowak}{2000}]{LN00}
{\sc Lazi\'c, R.} {\sc and} {\sc Nowak, D.} 2000.
\newblock A unifying approach to data independence.
\newblock In {\em Proceedings of the 11th International Conference on
  Concurrency Theory}. Lecture Notes in Computer Science, vol. 1877.
  Springer-Verlag, Pennsylvania, USA, 581--595.
\newblock \url*http://web.comlab.ox.ac.uk/oucl/research/areas/concurrency*.

\bibitem[\protect\citeauthoryear{McMillan}{McMillan}{1999}]{McM99}
{\sc McMillan, K.~L.} 1999.
\newblock Verification of infinite state systems by compositional model
  checking.
\newblock In {\em Conference on Correct Hardware Design and Verification
  Methods}. 219--234.

\bibitem[\protect\citeauthoryear{Namjoshi and Kurshan}{Namjoshi and
  Kurshan}{2000}]{NK00}
{\sc Namjoshi, K.} {\sc and} {\sc Kurshan, R.} 2000.
\newblock Syntactic program transformations for automatic abstraction.
\newblock In {\em Proceedings of the 12th International Conference on Computer
  Aided Verification}. Lecture Notes in Computer Science, vol. 1855.
  Springer-Verlag, 435--449.

\bibitem[\protect\citeauthoryear{Patterson and Hennessy}{Patterson and
  Hennessy}{1997}]{PH94}
{\sc Patterson, D.} {\sc and} {\sc Hennessy, J.} 1997.
\newblock {\em Computer Organization \& Design: The Hardware/Software
  Interface\/}, 2nd ed.
\newblock Morgan Kaufmann.

\bibitem[\protect\citeauthoryear{Qadeer}{Qadeer}{2001}]{Qad01}
{\sc Qadeer, S.} 2001.
\newblock Verifying sequential consistency on shared-memory multiprocessors by
  model checking.
\newblock Research Report 176, Compaq, Palo Alto, CA, USA.

\bibitem[\protect\citeauthoryear{Roscoe and Lazi\'{c}}{Roscoe and
  Lazi\'{c}}{2001}]{RL01}
{\sc Roscoe, A.} {\sc and} {\sc Lazi\'{c}, R.} 2001.
\newblock What can you decide about resetable arrays?
\newblock In {\em Proceedings of the 2nd International Workshop on Verification
  and Computational Logic ({VCL} 2001), Technical Report {DSSE-TR-2001-3},
  pages 5--23}. Declarative Systems and Software Engineering Research Group,
  Department of Electronics and Computer Science, University of Southampton,
  UK.

\bibitem[\protect\citeauthoryear{Stirling}{Stirling}{1992}]{Sti92}
{\sc Stirling, C.} 1992.
\newblock Modal and temporal logics.
\newblock In {\em Handbook of Logic in Computer Science}, {S.~Abramsky},
  {D.~Gabbay}, {and} {T.~Maibaum}, Eds. Vol.~2. Oxford University Press,
  477--563.

\bibitem[\protect\citeauthoryear{Wolper}{Wolper}{1986}]{Wol86}
{\sc Wolper, P.} 1986.
\newblock Expressing interesting properties of programs in propositional
  temporal logic.
\newblock In {\em Proceedings of the 13th {ACM} Symposium on Principles of
  Programming Languages}. 184--193.

\end{thebibliography}

\end{document}